\documentclass[preprint,12pt,authoryear]{elsarticle}

\usepackage{graphicx}
\usepackage{dcolumn}
\usepackage{bm}
\usepackage{xcolor}
\usepackage{booktabs}
\usepackage{tabularx}
\usepackage{array}

\usepackage{amssymb, amsmath, amsthm, mathrsfs}

\theoremstyle{plain}
\newtheorem{theorem}{Theorem}[section]

\newtheorem{proposition}[theorem]{Proposition}

\theoremstyle{definition}

\theoremstyle{remark}
\newtheorem{remark}{Remark}

\usepackage{hyperref}
\usepackage{cleveref}

\newcommand{\Qvec}{\mathbf{Q}}
\newcommand{\Mvec}{\mathbf{M}}
\newcommand{\pvec}{\mathbf{p}}
\newcommand{\nvec}{\mathbf{n}}
\newcommand{\mvec}{\mathbf{m}}

\graphicspath{{figures}}

\journal{Nuclear Physics B}

\begin{document}

\begin{frontmatter}

\title{Solution landscapes of ferronematics in microfluidic channels}

\author[inst1]{James Luke Dalby}
\author[inst2]{Apala Majumdar} 
\address[inst1]{Department of Mathematics, King’s College London, London WC2R 2LS, United Kingdom}
\address[inst2]{Department of Mathematics, University of Manchester, Manchester M13 9PL, United Kingdom}

\begin{abstract}
We study solution landscapes for ferronematics, i.e., a dilute suspension of magnetic nanoparticles in a nematic liquid crystal host, in a one-dimensional (1D) channel geometry. 
Adopting the modelling framework in \cite{bisht-2020-article}, the admissible ferronematic configurations are modelled in terms of critical points of an appropriately defined ferronematic energy. There are two types of critical points: full critical points that exploit all degrees of freedom and order reconstruction (OR) critical points. OR critical points have polydomain structures, with the polydomains separated by domain walls. We find that ferronematic systems are typically multistable, i.e., the free energy has multiple competing stable critical points, which model potentially physically observable configurations; the multistability is not a priori obvious. Further, we discover multiple stable and unstable OR solutions that differ in the locations and multiplicity of domain walls. The domain walls can act as binding sites in equilibrium processes and rafts for guided transport phenomena in non-equilibrium processes.
\end{abstract}


\end{frontmatter}

\section{Introduction}
Nematic liquid crystals (NLCs) are classical examples of partially ordered materials that are intermediate in character between the conventional solid and liquid phases of matter \cite{deGennes}. NLCs possess long-range orientational order, with their typically rod-shaped molecules aligned along locally distinguished/preferred directions known as ``directors" \cite{stewart}. This intrinsic directionality implies that NLCs readily respond to applied external fields (especially electric fields) making them useful materials for applications.
Most famously, NLCs  have been the working material of choice in the vast liquid crystal display industry \cite{lagerwallreview}. Unsurprisingly then, the mathematics of NLCs is a well developed field. As such, the next generation of NLC research and applications is upon us \cite{lagerwallreview}, and with this comes the need to study new NLC-type systems and their associated mathematical theories.

NLC applications typically exploit strong NLC responses to  external electric fields as opposed to magnetic fields, since the NLC response to magnetic fields is much weaker (perhaps seven orders of magnitude smaller) than their dielectric response \cite{stewart}.
First proposed by Brochard and de Gennes in the 1970s, 
ferronematics are a class of composite nematics comprising a suspension of magnetic nanoparticles (MNPs) in a nematic host that  subsequently exhibits a net nonzero 
magnetisation, even in the absence of external magnetic fields \cite{brochardgennes}.
Consequently, the magnetic susceptibility of the ferronematic system can be substantially increased (by several orders of 
magnitude) in comparison with pure NLC systems \cite{burylov} and magnetic switching can be efficiently achieved, making ferronematics an ideal candidate for the next generation of magnetically driven switchable NLC devices \cite{smalyukh}. 

Despite their early theoretical beginnings, ferronematics are relatively new and unexplored from an applications perspective, with the first stable ferronematic system reported in \cite{mertelj-2013-article} in 2013. Multistable ferronematic systems, i.e., systems that can support multiple stable ferronematic configurations acting as multiple modes of functionality, without an external magnetic field, are a concept of tremendous scientific interest. 
This is analogous to multistable NLC systems with multiple observable stable NLC configurations without any applied fields (e.g. the Zenithal Bistable Device \cite{ZBD}), and one can use an external electric field to switch between the competing stable configurations. Such multistable NLC systems have been used for display technologies, counterfeiting, etc. However, ferronematics have nematic and magnetic order that allows for greater complexity and richness in the observable equilibria \cite{dalby-farrell-majumdar-xia}.
From a theoretical perspective, one needs to understand both the complexity of the stable configurations and the pathways between the competing multiple stable configurations to model switchable soft NLC-based devices.  Solution landscapes are a key tool for this - computing the critical points, both stable and unstable, of a given free energy associated with a material system. The stable critical points (including energy minimisers) model the physically observable configurations whereas the unstable critical points play a critical role in the pathways between competing stable critical points and dictate the selection mechanisms for stable configurations in a multistable system  \cite{han_nonlinearity}. This paper is devoted to a study of solution landscapes of ferronematics in a 1D channel geometry, inspired by microfluidic systems (see \cite{dalby-farrell-majumdar-xia}, \cite{senugptaluxembourg}), with the purpose of predicting and controlling multistability in simplified ferronematic settings. 

We study a dilute suspension of MNPs, i.e., the size of the MNPs is small compared to the average distance between MNPs and the volume fraction of MNPs is much smaller than unity \cite{canevarizarnescu}, in a NLC-filled channel, in an one-dimensional (1D) setting. 
The MNPs generate a spontaneous magnetisation even in the absence of external magnetic fields and hence, the ferronematic mixture is described by two order parameters: a $\Qvec$-tensor for the nematic ordering and a magnetisation vector, $\Mvec$, to describe the spontaneous magnetisation induced by the suspended MNPs. The nematic order parameter contains information about the nematic directors and the degree of orientational ordering about them. The ferronematic free energy is a non-linear and non-convex functional of $\Qvec$ and $\Mvec$ and is the sum of three contributions: a nematic energy, a magnetisation/magnetic energy and a nemato-magnetic coupling energy \cite{dalby-farrell-majumdar-xia}. The nemato-magnetic coupling energy stems from the collective NLC-MNP interactions \cite{canevarizarnescu}. The rescaled ferronematic energy is parameterised by four dimensionless variables: two rescaled elastic constants, $l_1$ and $l_2$; a parameter $\xi$ that weighs the relative dominance of the nematic and magnetic energies and a NLC-MNP coupling parameter, $c$. The parameter $c$ depends on the shape of the MNPs and the boundary conditions on the MNP surfaces, which in turn dictate the NLC-MNP interactions.

This model has been considered previously in the same 1D setting in \cite{dalby-farrell-majumdar-xia,bisht-2019-article}, and a two-dimensional (2D) setting in \cite{bisht-2020-article, maity2021}.  
In \cite{bisht-2020-article}, the authors study ferronematic systems on square domains motivated by the planar bistable nematic device in \cite{tsakonas2007}; in \cite{han_ferronematics}, the authors study ferronematic configurations on regular 2D polygons while in \cite{dalby-farrell-majumdar-xia}, the authors prove analytic results on the existence, multiplicity and stability of critical points of the ferronematic free energy as proposed in  \cite{bisht-2020-article}, including an illuminating study of order reconstruction (OR) solutions. OR solutions are special because they are the union of nematic and magnetic polydomains. Polydomains are sub-domains with constant nematic directors and constant normalised magnetisation $\mvec = \frac{\Mvec}{|\Mvec|}$. The polydomains are separated by domain walls. Nematic domain walls are defined by $\Qvec=0$ in this reduced setting and magnetic domain walls are defined by $\Mvec=0$. The nematic director and $\mvec$ can jump across the domain walls and this jump is absorbed by the vanishing scalar order parameters, $|\Qvec|$ and $|\Mvec|$. In an uncoupled setting, OR solutions are simply solutions of the scalar Allen-Cahn equation and are well-studied. The NLC-MNP coupling introduces additional nonlinearities and proliferates the multiplicity of both stable and unstable OR solutions. The nematic and magnetic domain walls interact with each other and this interaction gives new scope to manipulate the domain walls in OR solutions, exploiting them for both equilibrium and non-equilibrium processes. In \cite{maity2021}, the authors conduct some elegant numerical analysis for the ferronematic model studied in the preceding papers.

Our work adds substantial value to previous work since we focus on solution landscapes of the ferronematic free energy in an albeit simplified channel geometry: solution landscapes include an exhaustive study of stable and unstable critical points of the ferronematic free energy and their Morse indices (not covered in previous work) as a function of the model parameters, accompanied by pathway maps that illustrate the connectivity of the critical points on the energy landscape (see \Cref{fig:illustration}). Hence, solution landscapes give crucial insight into both the multiplicity of stable ferronematic configurations and switching mechanisms between the competing stable ferronematic configurations. We compute the solution landscapes using the high-index optimisation-based shrinking dimer (HiOSD) method \cite{yin2019high} (see \Cref{sec:numerical_methods}). Informally, the Morse index is a measure of the degree of instability of a critical point so that the least unstable critical points have Morse index-$1$. In general, researchers focus on stable index-$0$ critical points that correspond to local or global minimisers of a free energy and model physically observable configurations. The index-$1$ saddle points, often referred to as transition states, typically connect competing stable critical points on transition pathways. However, switching can also be achieved by means of higher energy saddle points \cite{han_nonlinearity}. As has been shown in \cite{han_dalby_carter_majumdar_machon}, in some parameter regimes, the energy barrier (i.e., the energy difference between two critical points) for a pathway mediated by a higher-index saddle point can be close to that of a pathway mediated by an index-$1$ transition state, so that both pathways are physically relevant and likely to be observed in practice. 

We study two different types of critical points of the model ferronematic energy, subject to Dirichlet conditions for $\Qvec$ and $\Mvec$: full solutions without polydomains that exhibit  smooth reorientation of the nematic directors and $\Mvec$ throughout the channel geometry, and OR solutions which possess polydomains along with  nematic and magnetic domain walls. 
Such domain walls/disclination lines can be used in the architecture of micro-wires, or as soft rails for the transport of colloidal particles or droplets in microfluidic channels \cite{agha}. More generally, polydomain structures can have distinct optical and/or mechanical
properties if experimentally realised \cite{smalyukh}, making them interesting for applications. We demonstrate that (i) there can exist multiple stable OR solutions; (ii) OR solutions can act as transition states connecting pairs of stable full solutions and (iii) OR solutions can act as parent states in the solution landscape and connect any pair of stable and unstable critical points. OR solutions are likely to be stable for narrow channels whilst full solutions are stable for wide channels, although the stability criteria depend on a delicate balance between the phenomenological model parameters, $\left\{l_1, l_2, \xi, c \right\}$. The NLC-MNP coupling proliferates the multiplicity of full and OR solutions. Our study is exploratory and academic in nature, both in terms of the 1D setting and the choices of the model parameters. However, our work gives valuable insight into the complexity of ferronematic systems and how the relative weightings of energy terms affect the solution landscapes, i.e., what are the effects of decreasing $l_1$ or increasing $c$ and are there some universal underlying principles?  In the penultimate section of this paper, we discuss generalisations of our work to physically relevant parameter regimes, natural boundary conditions, and 2D analogues of our 1D study. 
Of course, there remain numerous open questions about physically pertinent parameters values for $l_1, l_2, \xi, c$; questions about how to experimentally tune these parameters, and the neglected physical effects in our model, e.g., out-of-plane magnetisation effects, other choices of the ferronematic free energy, etc.

The paper is organised as follows. In \Cref{sec:model}, we introduce the ferronematic problem and modelling framework. In \Cref{sec:numerical_methods}, we describe the HiOSD method.  In \Cref{sec:numerics}, we vary two model parameters, $l_1=l_2 := l$ and $c$ and compute the associated solution landscapes, to investigate how $l$ and $c$ can be tuned for bespoke multistable ferronematic systems. In \Cref{sec:real}, we discuss ferronematic solution landscapes with $l_2 < l_1$, natural boundary conditions for $\Mvec$ and 2D versions of our 1D study. In \Cref{sec:conclusion}, we present some conclusions and future directions.

\section{Model}\label{sec:model}

We work with a channel geometry as in \cite{dalby-farrell-majumdar-xia}, which is defined by $\tilde{\Omega}=\{(x,y,z)\in\mathbb{R}^3:-L\leq x\leq L,-D\leq y\leq D,0\leq z\leq H\}$, where $L \gg D$ and $D\gg H$. $L$ is half the channel length, $D$ is half the channel width and $H$ is the channel height. 
We assume planar surface anchoring conditions on the top and bottom channel surfaces at $z=0$ and $z=H$, which means that the NLC molecules lie in the plane on these surfaces, i.e., the $xy$-plane without a specified direction. 
We further impose periodic boundary conditions on $x=\pm L$ and Dirichlet boundary conditions on $y=\pm D$. 
Since $H \ll D \ll L$, we assume that the structural properties are invariant in the $z$-direction and invariant in the $x$-direction; then it suffices to consider a 1D channel geometry $\Omega = [-D, D]$.  

We model a dilute suspension of ferronematics inside the 1D channel as described above. In other words, we model a suspension of MNPs inside a NLC-filled channel, such that the size of the MNPs ($a$) is much smaller than the distance between the MNPs ($d$), which in turn is much smaller than the channel dimensions so that $a \ll d \ll D$. The MNPs are well dispersed, do not interact with each other and induce a spontaneous magnetisation. The interaction of the MNPs with the ambient NLC is described in terms of a homogenised coupling energy as described below. 
Following the modelling framework in \cite{dalby-farrell-majumdar-xia,bisht-2019-article,bisht-2020-article,maity2021,han_ferronematics}, we use a reduced Landau-de Gennes (rLdG) $\Qvec$-tensor to describe the nematic host and the spontaneous magnetisation is modelled by a 2D vector, $\Mvec=(M_1,M_2)\in\mathbb{R}^2$. 

The rLdG $\Qvec$-tensor stems from a rigorous dimension reduction argument from three dimensions. In the conventional LdG theory, the nematic phase is modelled by the LdG $\Qvec$-tensor order parameter, which is a symmetric and traceless $3\times 3$ matrix, with five degrees of freedom. Following Theorem 5.1 in \cite{golovaty2015} (also see Theorem 2.1 in \cite{wang_canevari_majumdar_2019}) which applies to thin microfluidic channels with boundary conditions such as ours, one can deduce that the physically relevant (energy minimising) LdG $\Qvec$-tensors are $z$-invariant, have a fixed eigenvector $\hat{z}$ (the unit-vector in the $z$-direction) with associated fixed eigenvalue. This necessarily reduces the number of degrees of freedom from five in the full framework, to two degrees of freedom in the rLdG framework.

We denote the full LdG $\Qvec$-tensor by $\Qvec_f$, and the rLdG $\Qvec$-tensor by $\Qvec_{2D}$ below. 
\begin{equation} \label{eq:3dextrapolation}
    \Qvec_f (x,y,z)=
            \begin{pmatrix}
            -q_3 & 0 & 0\\
            0 & -q_3 & 0\\
            0 & 0 & 2q_3
        \end{pmatrix} + \begin{pmatrix}
            Q_{11}(x,y) & Q_{12}(x,y) & 0\\
            Q_{12}(x,y) & -Q_{11}(x,y) & 0\\
            0 & 0 & 0
        \end{pmatrix},
\end{equation} such that
\[
\Qvec_{2D} = \begin{pmatrix} Q_{11} & Q_{12} \\
            Q_{12} & -Q_{11}
            \end{pmatrix}
\] and $q_3$ is a fixed known constant determined by the temperature and boundary conditions on $z=0, H$.
In what follows, we model the nematic phase by the rLdG $\Qvec_{2D}$-tensor and drop the subscript $2D$ from $\Qvec_{2D}$.
More precisely, the rLdG $\Qvec$-tensor is a symmetric, traceless $2\times 2$ matrix that can be equivalently expressed as:
\begin{equation}
    \Qvec=s\left(2\nvec\otimes\nvec-\mathbf{I}_2\right),\label{eq:Q}
\end{equation}
where $\nvec = \left(\cos \vartheta, \sin \vartheta \right)\in\mathbb{S}^1$ is the nematic director in the $xy$-plane, $\vartheta$ is the angle between $\nvec$ and the $x$-axis, $s$ is a scalar order parameter which measures the degree of nematic ordering about $\nvec$, 
and $\mathbf{I}_2$ is the $2\times 2$ identity matrix. The nematic director can be interpreted as the preferred direction of averaged molecular alignment in the $xy$-plane. 
Then the two independent components, $Q_{11}$ and $Q_{12}$ are related to $s$ and $\vartheta$ by
\begin{equation}
    Q_{11} = s\cos 2 \vartheta,\quad Q_{12} = s \sin 2\vartheta.
\end{equation}
        
In what follows, we take our domain to be $\Omega=\left\{ -D \leq y \leq D \right\}$, assume that $\Qvec$ and $\Mvec$ only depend on $y$ and work in the admissible space
\[
\Qvec \in W^{1,2}\left(\Omega; S_0 \right); ~ \Mvec \in W^{1,2}\left(\Omega; \mathbb{R}^2 \right)
\] where $W^{1,2}$ is the standard Sobolev space of measurable and square-integrable functions with square-integrable first derivatives \cite{sobolev}. This 1D profile can be extrapolated to three dimensions as above:
\[
\Qvec_f (x,y, z)=
            \begin{pmatrix}
            -q_3 & 0 & 0\\
            0 & -q_3 & 0\\
            0 & 0 & 2q_3
        \end{pmatrix} + \begin{pmatrix}
            Q_{11}(y) & Q_{12}(y) & 0\\
            Q_{12}(y) & -Q_{11}(y) & 0\\
            0 & 0 & 0
        \end{pmatrix}.
\]

We model the physically observable $\Qvec$ and $\Mvec$ profiles as minimisers of an appropriately defined ferronematic free energy as in \cite{dalby-farrell-majumdar-xia}. The ferronematic free energy density is the sum of three contributions: a nematic energy density, a magnetic energy density and a nemato-magnetic coupling energy density: 
\begin{equation}
    f_{nem}=\frac{K}{2}|\nabla\Qvec|^2 + 2A|\Qvec|^2 
    + C|\Qvec|^4 + \frac{|A|^2}{C}, \label{eq:f_nem}
\end{equation}
\begin{equation}
    f_{mag} = \frac{\kappa}{2} |\nabla \Mvec|^2 + \frac{\alpha}{2}|\Mvec|^2
    + \frac{\beta}{4}|\Mvec|^4+\frac{|\alpha|^2}{4\beta}, \label{eq:f_mag}
\end{equation}
\begin{equation}
    f_{coupling} = -\gamma \mu_0 \Mvec ^T \Qvec \Mvec = - \gamma\mu_0 s |\Mvec|^2\left(2(\nvec\cdot\mvec)^2 - 1 \right). \label{eq:f_couple}
\end{equation}
$K$ is a positive nematic elastic constant, $A = a_0 (T - T^*)$ where $a_0$ and $T^*$ are material-dependent constants and $T$ is the temperature, and $C>0$ is a material dependent positive constant. We work with a fixed temperature $T < T^*$ so that $A<0$, and ordered nematic phases with $\Qvec \neq 0$ are energetically favourable. The nematic energy, \eqref{eq:f_nem} is the standard reduced LdG free energy as used in \cite{han_nonlinearity}. The magnetic energy, \eqref{eq:f_mag} is a Ginzburg-Landau energy, where $\alpha<0$ and $\beta>0$ are Landau coefficients describing the ferronematic transition \cite{mertelj-2013-article}, and $\kappa$ is a phenomenological stiffness constant that penalises spatial inhomogeneities in $\Mvec$ \cite{bisht-2020-article}. 
Finally, the coupling energy \eqref{eq:f_couple} can be derived from rigorous homogenisation principles for a dilute suspension of MNPs in a nematic host; the coupling constant $\gamma$ contains information about the shape of the MNPs and the NLC-MNP interactions \cite{canevarizarnescu}. We treat $\gamma$ as a material-dependent fixed constant; positive values of $\gamma$ coerce $\nvec$ and $\mvec = \frac{\Mvec}{|\Mvec|}$ to be parallel or anti-parallel whereas negative values of $\gamma$ coerce $\nvec$ and $\mvec$ to be
perpendicular to each other, when $|\Mvec| \neq 0$  ($\mu_0$ is the vacuum permeability \cite{mertelj-2013-article}).

We use the following scalings, $\bar{\Qvec}=\sqrt{\frac{2C}{|A|}}\Qvec$, $\bar{\Mvec}=\sqrt{\frac{\beta}{|\alpha|}}\Mvec$ and $\bar{y}=y/D$, which leads to four dimensionless parameters:
\begin{multline}
        \label{eq:dimenless-def}
        l_1=\frac{K}{D^2|A|},\;l_2=\frac{\kappa}{D^2|\alpha|},\; c=\frac{\gamma\mu_0}{|A|}\sqrt{\frac{C}{2|A|}}\frac{|\alpha|}{\beta},\;\xi=\frac{C}{|A|^2}\frac{|\alpha|^2}{\beta}.
\end{multline}
Henceforth we drop bars for convenience and work with dimensionless variables.
The rescaled domain is $\Omega = [-1, 1]$ and the total rescaled dimensionless ferronematic free energy for this 1D problem is given by \cite{dalby-farrell-majumdar-xia}:
\begin{equation}
    \label{eq:energy}
    \begin{aligned}
        F&(Q_{11}, Q_{12}, M_1, M_2)
    := \\
    &\int^1_{-1} \Bigg\{\frac{l_1}{2}\left[\left( \frac{\mathrm{d} Q_{11}}{\mathrm{d} y}\right)^2 + \left(\frac{\mathrm{d} Q_{12}}{\mathrm{d} y}\right)^2 \right] + \left( Q_{11}^2 + Q_{12}^2 - 1 \right)^2 \\
    & + \frac{\xi l_2}{2} \left[ \left(\frac{\mathrm{d} M_1}{\mathrm{d} y}\right)^2 + \left(\frac{\mathrm{d} M_2}{\mathrm{d} y}\right)^2 \right] + \frac{\xi}{4}\left(M_1^2 + M_2^2 - 1 \right)^2 \\
    & - cQ_{11}\left(M_1^2 - M_2^2 \right) - 2c Q_{12}M_1 M_2 \Bigg\} ~\mathrm{d}y.
    \end{aligned}
\end{equation}
We interpret $l_1>0$ and $l_2>0$ as scaled elastic constants (inversely proportional to $D^2$, i.e., half the channel width squared), $\xi$ weighs the relative strength of the nematic and magnetic energies and $c$ is a coupling parameter. 
The sign of $c$ is controlled by the sign of $\gamma$; positive $c$ promotes co-alignment of $\nvec$ and $\mvec$. 

We work with a fixed $A<0$ and there is no uniform consensus on the relative values of the four material-dependent and geometry-dependent coefficients: $\left\{l_1, l_2, \xi, c \right\}$, since they are tuneable parameters.
We make some comments on physically relevant values for our dimensionless parameters.  Consider  $l_1=\frac{K}{|A|D^2}$; from \cite{Zumar2009}, some typical values for the liquid crystal 5CB are $K=4\times10^{-11}$N and $A=-0.172\times 10^6$Nm$^{-2}$; 
combined with a channel width of $D=0.5\times 10^{-5}$m, this yields a characteristic value of $l_1=9.3\times10^{-6}$. The values of $\xi$ and $l_2$ could be small for dilute ferronematic suspensions, since the magnetic energy is  dominated by the nematic energy in the dilute regime. One can interpret $\xi$ as being the ratio of the minimisers of the LdG bulk energy and the magnetic bulk energy without coupling i.e.
\[
\min 2A|\Qvec|^2 
    + C|\Qvec|^4 = -\frac{A^2}{C} ; \min \frac{\alpha}{2}|\Mvec|^2
    + \frac{\beta}{4}|\Mvec|^4 = -\frac{\alpha^2}{4 \beta}.
\]

From \cite{ZBD18b}, one can calculate typical values of $\frac{\alpha}{\beta}$, $\alpha$ and can estimate $\xi \in \left(10^{-6}, 10^{-5} \right)$. Similarly, the coupling constant $c$ can also be estimated by the values of $\mu_0$, $\frac{\alpha}{\beta}$ and $A$, and typical estimates are $c \approx 10^{-6}$. In this respect, one can take $l_1, \xi, c$ to be of comparable magnitude. We expect $\xi l_2 \ll l_1$ for a dilute suspension but cannot estimate $l_2$ explicitly in the absence of physical values for $\kappa$ as such.  

The minimisers and (in fact, all critical points) of the free energy \eqref{eq:energy} are classical solutions of the corresponding Euler-Lagrange equations \cite{dalby-farrell-majumdar-xia}:
\begin{subequations}
    \label{eq:euler-lagrange}
\begin{align}
    & l_1 \frac{\mathrm{d}^2 Q_{11}}{\mathrm{d}y^2} = 4 Q_{11}(Q_{11}^2 + Q_{12}^2 - 1) - c\left(M_1^2 - M_2^2 \right),\label{eq:Q11_again} \\
    & l_1 \frac{\mathrm{d}^2 Q_{12}}{\mathrm{d}y^2} = 4 Q_{12}(Q_{11}^2 + Q_{12}^2 - 1) - 2cM_1 M_2,\label{eq:Q12_again}\\
    & \xi l_2 \frac{\mathrm{d}^2 M_{1}}{\mathrm{d}y^2} = \xi M_1 \left(M_1^2 + M_2^2 - 1 \right) - 2c Q_{11}M_1 - 2c Q_{12}M_2,\label{eq:M1_again} \\
    & \xi l_2 \frac{\mathrm{d}^2 M_{2}}{\mathrm{d}y^2} = \xi M_2 \left(M_1^2 + M_2^2 - 1 \right) + 2c Q_{11} M_2 - 2c Q_{12} M_1\label{eq:M2_again}.
\end{align}
\end{subequations}
In what follows, we assume $l_1 = l_2 = l >0$, fix $\xi=1$ and vary $l_1$ and $c$ in the ranges, $l_1 \in [0.1, 1]$ and $c \in [1,5]$. 
We also perform some numerical experiments with $l_2 = 0.01 l_1$ to explore the effects of small $l_2$ on ferronematic solution landscapes.
We solve this system subject to Dirichlet conditions for $\Qvec$ and $\Mvec$ on the boundaries $y=\pm 1$, summarised in \Cref{table:BCs-ferronematics}. 
\begin{table}[htbp]
\centering
\caption{Dirichlet conditions for $\Qvec$ and $\Mvec$.}
\label{table:BCs-ferronematics}
\begin{tabular}{|c|c|c|c|c|c|c|}
\hline
Edge & $Q_{11}$ & $Q_{12}$ & $\nvec$ & $s$ & $M_1$ & $M_2$ \\
\hline
$y=-1$ & $1$ & $0$ & $(1,0)$ & $1$ & $1$ & $0$ \\
\hline
$y=+1$ & $-1$ & $0$ & $(0,1)$ & $1$ & $-1$ & $0$ \\
\hline
\end{tabular}
\end{table}

We impose mixed mutually orthogonal boundary conditions for $\nvec$, i.e., $\nvec = (1, 0)$ on $y=-1$ and $\nvec = (0,1)$ on $y=1$ ($\nvec$ is only determined up to a sign). The boundary conditions for $\Mvec$ describe a $\pi$-rotation between the bounding plates, $y=\pm 1$. These boundary conditions have been chosen to match those in \cite{bisht-2019-article,dalby-farrell-majumdar-xia} so that we can build on our understanding from these papers. Different boundary conditions would inevitably lead to different conclusions. 
Further, Dirichlet boundary conditions are commonly used for modelling nematic systems, since experimentalists can prescribe nematic alignment on geometric boundaries with some confidence; see work on Freedericksz transitions in channel geometries in \cite{freedericksz} and the planar bistable nematic device \cite{tsakonas2007}. We choose to work with Dirichlet conditions for $\Mvec$ but there is no clear consensus on how $\Mvec$ can be constrained on the channel edges. We largely work with $l_1 = l_2$ and in Section~\ref{sec:real}, we numerically compute critical points of the ferronematic energy with $l_2 = 0.01 l_1$, natural boundary conditions for $\Mvec$ accompanied by some 2D studies for which $\Qvec$ and $\Mvec$ are allowed to vary in both the $x$ and $y$-directions of the channel cross-section. These numerical generalisations are not exhaustive but help us test the robustness of our simplified 1D study in wider settings.

\begin{remark}\label{rem:symmetry}
There are two distinct and important solution types for the system of ordinary differential equations in \eqref{eq:euler-lagrange}. Firstly, we have full solutions of the form $(\Qvec,\Mvec)=(Q_{11},Q_{12},M_1,M_2)$, which exploit all four degrees of freedom. From the Euler-Lagrange equations \eqref{eq:euler-lagrange}, it is clear that if $\Qvec_1=(\tilde{Q}_{11},\tilde{Q}_{12}, \tilde{M}_1,\tilde{M}_2)$ is a solution, then $\Qvec_2 = (\tilde{Q}_{11},-\tilde{Q}_{12}, \tilde{M}_1,-\tilde{M}_2)$ is also a solution of \eqref{eq:euler-lagrange} and the boundary conditions in Table~\ref{table:BCs-ferronematics} are satisfied. All full solutions come in pairs which have the same $Q_{11}$ and $M_1$ profiles, whilst the $Q_{12}$ and $M_2$ profiles are reflections of each other. From \eqref{eq:energy},  $\Qvec_1$ and $\Qvec_2$ are energetically degenerate. There are more admissible symmetry groups for the solution pairs with natural boundary conditions for $\Mvec$, as will be evident in Section~\ref{sec:real}.

The second solution type are OR solutions of the form $(\Qvec,\Mvec)=(Q_{11},0,M_1,0)$, i.e., $Q_{12}=M_2\equiv 0$ throughout $\Omega$ (such a solution branch is compatible with \eqref{eq:Q12_again} and \eqref{eq:M2_again} and the boundary conditions in Table~\ref{table:BCs-ferronematics}, $\forall l,c>0$). OR solutions only have two degrees of freedom; further
\[
Q_{12} \equiv 0 \implies \nvec \in \left\{(1,0), (-1,0), (0,1), (0,-1) \right\} \quad \textrm{if $s \neq 0$}
\]
since $Q_{12} = s \sin 2\vartheta$ when $\nvec = (\cos\vartheta, \sin \vartheta)$. Hence, the nematic director is constrained to belong to the discrete set above and $\mvec = \frac{\Mvec}{|\Mvec|} \in \left\{(1,0), (-1,0) \right\}$. Thus, the OR solution is composed of nematic and magnetic polydomains. The nematic  polydomains have constant $\nvec$ (where $\nvec$ can take one of the four values above) and the magnetic polydomains have constant $\mvec$, where $\mvec=\left(\pm 1, 0 \right)$. The polydomains are separated by domain walls; the nematic domain walls are labelled by $\Qvec=0$  and the magnetic domain walls are points with $\Mvec =M_{1}=M_{2}=0$. The points with $\Qvec=0$ and $\Mvec=0$ extend to lines in 2D and to domain walls in three dimensions (3D). We note that such points must exist for OR solutions because of the conflicting boundary conditions for $Q_{11}$ and $M_1$ in Table~\ref{table:BCs-ferronematics}. Further, the nematic and magnetic domain walls need not coincide and indeed, there is ample flexibility in both the locations and multiplicity of nematic and magnetic domain walls, subject to the Dirichlet conditions above. 
\end{remark}

\section{Numerical scheme}\label{sec:numerical_methods}

We employ the HiOSD method as developed in \cite{yin2019high, yin2020construction}, to numerically compute both stable and unstable solutions of \eqref{eq:euler-lagrange} (or equivalently, minimisers and saddle points of the ferronematic energy in \eqref{eq:energy}). The code base to perform simulations is built by hand in Python. This method is described in detail below.

\subsection{Critical points}

A critical point $(\Qvec,\Mvec)$ of \eqref{eq:energy} is classified as stable if the corresponding Hessian of the free energy $F(\Qvec,\Mvec)$ is positive definite, or equivalently, if all its eigenvalues are positive, and is unstable otherwise. Here and hereafter, we will denote a critical point of \eqref{eq:energy} by
\begin{equation}
    (\Qvec,\Mvec):=(Q_{11},Q_{12},M_1,M_2),
\end{equation}
i.e., a single vector containing information about the $\Qvec$ and $\Mvec$ profiles, in this order. We solve the time-independent Euler-Lagrange equations \eqref{eq:euler-lagrange} via the gradient flow equations:
\begin{subequations}
    \label{eq:gradient_flow_equations}
\begin{align}
    & \frac{\partial Q_{11}}{\partial t}=l_1 \frac{\partial^2 Q_{11}}{\partial y^2} - 4 Q_{11}(Q_{11}^2 + Q_{12}^2 - 1) + c\left(M_1^2 - M_2^2 \right),\label{eq:Q11-flow} \\
    & \frac{\partial Q_{12}}{\partial t}=l_1 \frac{\partial^2 Q_{12}}{\partial y^2} - 4 Q_{12}(Q_{11}^2 + Q_{12}^2 - 1) + 2cM_1 M_2,\label{eq:Q12-flow}\\
    & \frac{\partial M_{1}}{\partial t} = \xi l_2 \frac{\partial^2 M_{1}}{\partial y^2} - \xi M_1 \left(M_1^2 + M_2^2 - 1 \right) \nonumber\\
&\quad\quad\quad\quad\quad\quad\quad\quad\quad\quad\quad\quad + 2c Q_{11}M_1+ 2c Q_{12}M_2,\label{eq:M1-flow} \\
    & \frac{\partial M_{2}}{\partial t} = \xi l_2 \frac{\partial^2 M_{2}}{\partial y^2} - \xi M_2 \left(M_1^2 + M_2^2 - 1 \right) \nonumber\\
&\quad\quad\quad\quad\quad\quad\quad\quad\quad\quad\quad\quad - 2c Q_{11} M_2 + 2c Q_{12} M_1\label{eq:M2-flow}.
\end{align}
\end{subequations}
There is no inertia or hydrodynamics (since we neglect nematic flow and focus on equilibrium properties, as opposed to non-equilibrium properties) in this formulation and we set the dissipation constant to be unity; the time is fictitious and the gradient flow dynamics simply mimic relaxation to a critical point of the free energy with suitable initial conditions, along a path of decreasing energy on the energy landscape \cite{yin2019high, yin2020construction}. The derivatives are approximated using finite difference methods on a uniform mesh with mesh size $1/50$ partitioning $[-1,1]$. The time step update is governed by the HiOSD method described next.

The method is deemed to have converged to a time-independent solution of \eqref{eq:gradient_flow_equations} or a critical point of the free energy \eqref{eq:energy}, when the norm of the gradient of $F$ (denoted by $\nabla F$) has fallen below $10^{-6}$, i.e.,
\begin{equation}
    ||\nabla F(\Qvec,\Mvec)||_{L^2}<10^{-6},
\end{equation}
where
\begin{multline*}
    \nabla F(\Qvec,\Mvec)=\\
    \begin{pmatrix}
    l_1 \frac{\mathrm{d}^2 Q_{11}}{\mathrm{d} y^2} - 4 Q_{11}(Q_{11}^2 + Q_{12}^2 - 1) + c\left(M_1^2 - M_2^2 \right)\\
    l_1 \frac{\mathrm{d}^2 Q_{12}}{\mathrm{d}y^2} - 4 Q_{12}(Q_{11}^2 + Q_{12}^2 - 1) + 2cM_1 M_2\\
    \xi l_2 \frac{\mathrm{d}^2 M_{1}}{\mathrm{d}y^2} - \xi M_1 \left(M_1^2 + M_2^2 - 1 \right) + 2c Q_{11}M_1 + 2c Q_{12}M_2\\
    \xi l_2 \frac{\mathrm{d}^2 M_{2}}{\mathrm{d}y^2} - \xi M_2 \left(M_1^2 + M_2^2 - 1 \right) - 2c Q_{11} M_2 + 2c Q_{12} M_1
    \end{pmatrix}.
\end{multline*}

\subsection{The HiOSD method}

We compute solution landscapes of the ferronematic system and the Morse index of critical points of \eqref{eq:energy} using the HiOSD method \cite{yin2019high}. The Morse index of any critical point is the number of negative eigenvalues of the Hessian of the free energy \cite{milnor1963morse}. Hence, a critical point is unstable if it has Morse index greater or equal to 1 (i.e, it has at least one negative eigenvalue of its Hessian). An index-$k$ saddle point is an unstable critical point which has Morse index-$k$ and hence $k$ negative eigenvalues of the associated Hessian, making it unstable in $k$-distinguished eigendirections and stable in all other directions. By definition, these index-$k$ saddles are non-energy minimising critical points of \eqref{eq:energy}.

\begin{figure}
\centering
    \begin{minipage}{0.42\textwidth}
     \centering
        \includegraphics[width=1.0\textwidth]{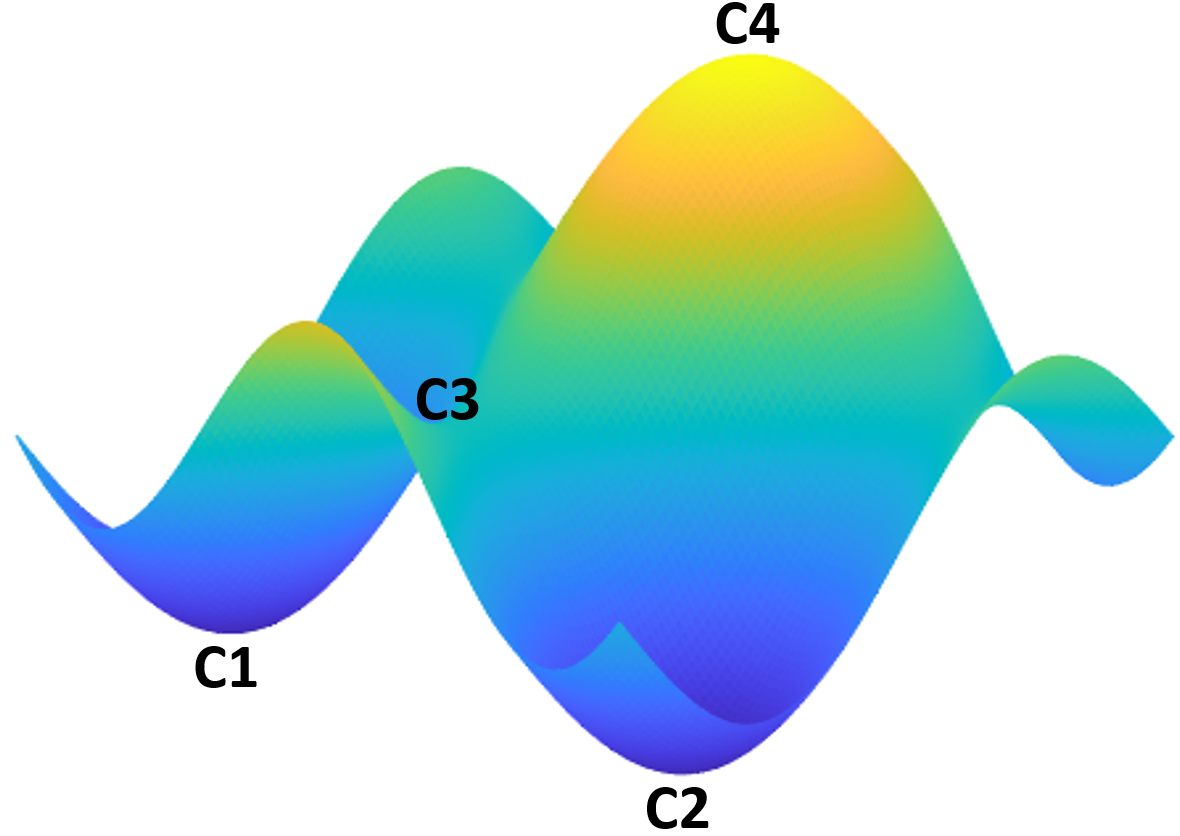}\\
        \ \textrm{Energy landscape}
    \end{minipage}
    \begin{minipage}{0.36\textwidth}
        \centering
        \includegraphics[width=0.7\textwidth]{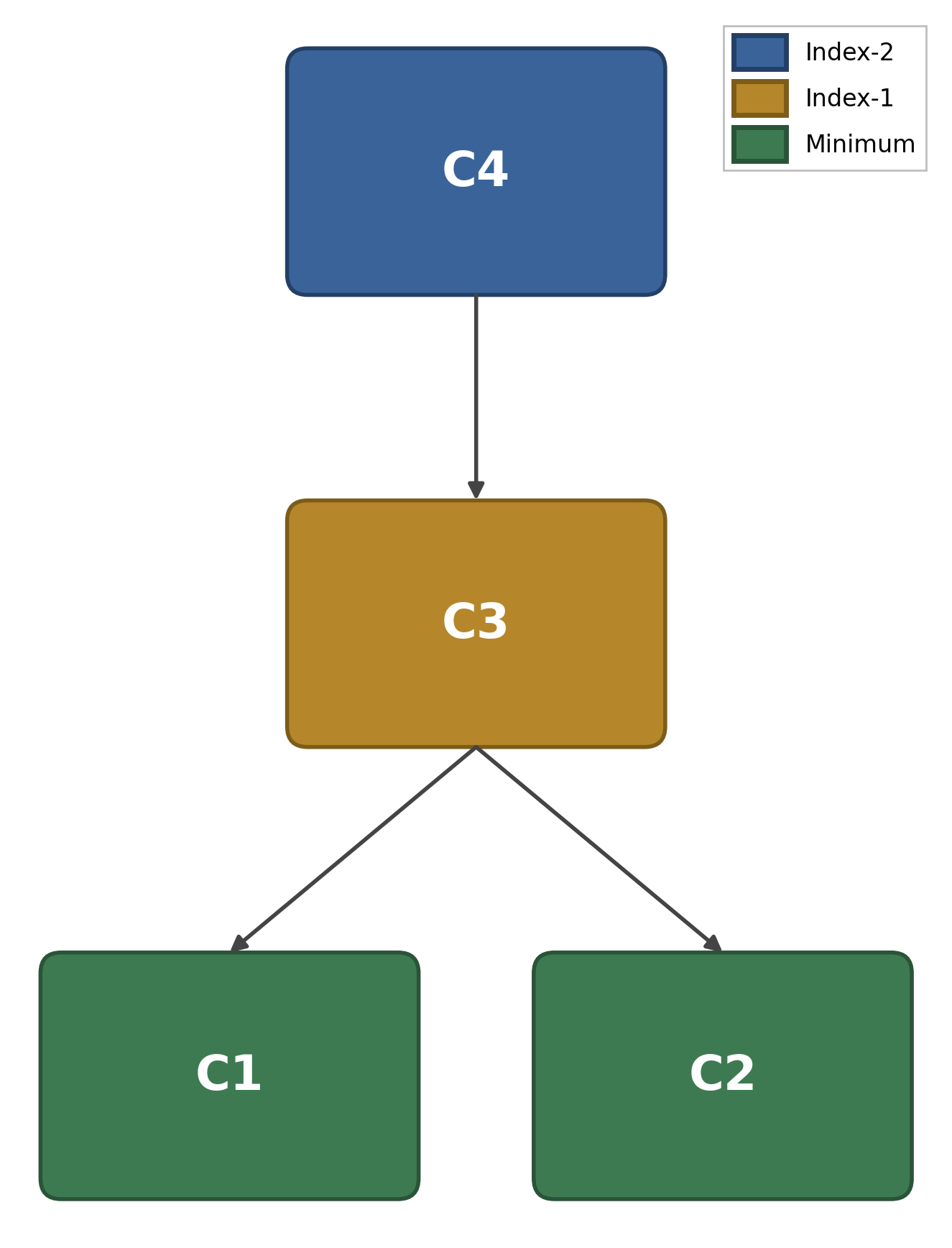}\\
        \ \textrm{Solution landscape}
    \end{minipage}
    \caption{A sketch of an energy landscape and its associated solution landscape (\cite{yin2020construction}). The $C_i$ denote critical points of the energy functional.} 
    \label{fig:illustration}
\end{figure}

The HiOSD method is a local-search algorithm for the computation of saddle points of arbitrary Morse index, which can be viewed as a generalisation of the optimisation-based shrinking dimer method for finding index-$1$ saddle points \cite{zhang2016optimization}. 
We combine the HiOSD method \cite{yin2019high} with upward/downward search algorithms \cite{yin2020construction} to compute solution landscapes, or equivalently, pathway maps for this ferronematic problem. We say that there exists a pathway between two critical points $(\Qvec_1,\Mvec_1)$ and $(\Qvec_2,\Mvec_2)$ of \eqref{eq:energy}, if we can follow a stable or unstable eigendirection (of the Hessian) of $(\Qvec_1,\Mvec_1)$ in the HiOSD method and find $(\Qvec_2,\Mvec_2)$ \cite{yin2020construction}. To make these ideas clear, an example solution landscape and associated energy landscape are presented in \Cref{fig:illustration}.

For a non-degenerate index-$k$ saddle point $(\hat{\mathbf{Q}},\hat{\Mvec})$, the Hessian $\mathbb{H}(\mathbf{Q},\Mvec)=\nabla^2 F(\mathbf{Q},\Mvec)$ at $(\hat{\mathbf{Q}},\hat{\Mvec})$, has exactly $k$ negative eigenvalues $\hat{\lambda}_1\leqslant \cdots \leqslant\hat{\lambda}_k$ with corresponding unit eigenvectors $\hat{\mathbf{v}}_1,\ldots, \hat{\mathbf{v}}_k$, satisfying $\big\langle\hat{\mathbf{v}}_j, \hat{\mathbf{v}}_i\big\rangle = \delta_{ij}$, $1\leqslant i, j \leqslant k$.
We define a $k$-dimensional subspace $\hat{\mathcal{V}}=\mathrm{span}\big\{\hat{\mathbf{v}}_1,\ldots, \hat{\mathbf{v}}_k\big\}$, then $(\hat {\mathbf{Q}},\hat{\Mvec})$ is a local maximum on a $k$-dimensional linear manifold $(\hat{\mathbf{Q}},\hat{\Mvec})+\hat{\mathcal{V}}$ and a local minimum on $(\hat{\mathbf{Q}},\hat{\Mvec})+\hat{\mathcal{V}}^\perp$, where $\hat{\mathcal{V}}^\perp$ is the orthogonal complement space of $\hat{\mathcal{V}}$.

The index-$k$ saddle $(\hat{\mathbf{Q}},\hat{\Mvec})$ is computed by means of a minimax optimisation problem
\begin{equation}
\min_{(\mathbf{Q},\Mvec)_{\hat{\mathcal{V}}^{\perp}}\in\hat{\mathcal{V}}^{\perp}}\max_{(\mathbf{Q},\Mvec)_{\hat{\mathcal{V}}}\in\hat{\mathcal{V}}} F((\mathbf{Q},\Mvec)_{\hat{\mathcal{V}}^{\perp}} + (\mathbf{Q},\Mvec)_{\hat{\mathcal{V}}}),\label{eq:minimax}
\end{equation}
where $(\mathbf{Q},\Mvec)_{\hat{\mathcal{V}}} = \mathcal{P}_{\hat{\mathcal{V}}}(\mathbf{Q},\Mvec)$ is the orthogonal projection of $(\mathbf{Q},\Mvec)$ on $\hat{\mathcal{V}}$, and $(\mathbf{Q},\Mvec)_{\hat{\mathcal{V}}^{\perp}} = (\mathbf{Q},\Mvec) - \mathcal{P}_{\hat{\mathcal{V}}}(\mathbf{Q},\Mvec)$. Here, $(\Qvec,\Mvec)$ is assumed to be sufficiently close to $(\hat{\Qvec},\hat{\Mvec})$ so that $\mathbb{H}(\Qvec,\Mvec)$ also has exactly $k$ negative eigenvalues, with corresponding eigenvectors $\mathbf{v}_i$, for $i=1,\ldots k$. Furthermore, we take $\mathcal{V}=\mathrm{span}\big\{\mathbf{v}_1,\ldots, \mathbf{v}_k\big\}$ to approximate $\hat{\mathcal{V}}$. 

For brevity of notation, in what follows $^.$ will denote differentiation with respect to $t$. To find a solution of the minimax problem \eqref{eq:minimax}, the dynamics of $(\Qvec,\Mvec)$ should satisfy that $\mathcal{P}_{\mathcal{V}}(\dot{\Qvec},\dot{\Mvec})$ is an ascent direction on the subspace $\mathcal{V}$ and that $(\dot{\Qvec},\dot{\Mvec})-\mathcal{P}_{\mathcal{V}}(\dot{\Qvec},\dot{\Mvec})$ is a descent direction on the subspace $\mathcal{V}^\perp$. This ensures $(\Qvec,\Mvec)$ increases to a maximum on $\mathcal{V}$ and decreases to a minimum on $\mathcal{V}^\perp$. Therefore, we take $\mathcal{P}_\mathcal{V}\nabla F(\Qvec,\Mvec)$ as the direction of $\mathcal{P}_\mathcal{V}(\dot{\Qvec},\dot{\Mvec})$ and $-\nabla F(\Qvec,\Mvec)+\mathcal{P}_\mathcal{V}\nabla F(\Qvec,\Mvec)$ as the direction of $(\dot{\Qvec},\dot{\Mvec})-\mathcal{P}_{\mathcal{V}}(\dot{\Qvec},\dot{\Mvec})$. The following gradient dynamics can then be used to solve \eqref{eq:minimax}
\begin{equation}
    \eta_1^{-1}(\dot{\Qvec},\dot{\Mvec})=-\nabla F(\Qvec,\Mvec)+2\mathcal{P}_{\mathcal{V}}\nabla F(\Qvec,\Mvec),
\end{equation}
where $\eta_1>0$ is a relaxation parameter.

The $k$-dimensional subspace $\mathcal{V}$ is constructed by solving the following constrained optimisation problem for the $i$'th eigenvector $\mathbf{v}_i$ (where $i=1,\ldots,k$) 
\begin{equation}\label{eq:quotient_problem}
\min_{\mathbf{v}_i\in\mathbb{R}^n} \;\langle \mathbb{H}(\mathbf{Q},\Mvec)\mathbf{v}_i, \mathbf{v}_i\rangle,\;
\mathrm{s.t.} \;\langle \mathbf{v}_j,\mathbf{v}_i\rangle=\delta_{ij},\; j=1,2,\ldots,i.
\end{equation}
Using the simultaneous Rayleigh-quotient iterative minimisation method \cite{rayleigh} to solve \eqref{eq:quotient_problem}, and noting that under the orthonormal constraint, $\mathcal{P}_{\mathcal{V}}$ is given by $\Sigma_{i=1}^k \mathbf{v}_i \mathbf{v}_i^\top$, we obtain a dynamical system of the HiOSD for finding an index-$k$ saddle point, 
\begin{equation}\label{eq:dynamics}
\left\{
\begin{aligned}
&\eta_1^{-1}(\dot{\mathbf{Q}},\dot{\Mvec})    =- \left(\mathbf{I}-2\sum_{j=1}^{k}\mathbf{v}_j \mathbf{v}_j^\top\right)\nabla F(\mathbf{Q},\Mvec), \\&
\eta_2^{-1}\dot{\mathbf{v}}_i  = -\left(\mathbf{I}-\mathbf{v}_i\mathbf{v}_i^\top-2\sum_{j=1}^{i-1}\mathbf{v}_j \mathbf{v}_j^\top\right)\mathbb{H}(\mathbf{Q},\Mvec)\mathbf{v}_i,\; \\
&\qquad\qquad\qquad\qquad\qquad\qquad\qquad\qquad\textrm{for } i=1,\ldots,k,
\end{aligned}
\right.
 \end{equation}
where the state variable $(\mathbf{Q},\Mvec)$ and $k$ direction variables $\mathbf{v}_i$ are coupled, $\mathbf{I}$ is the identity operator and $\eta_2>0$ is a relaxation parameter.

We implement Algorithm 1 in \cite{yin2019high} with relaxation parameters $\eta_1=\eta_2=\Delta t$, where $\Delta t$ is our Euler time step size used to solve the gradient flow equations \eqref{eq:Q11-flow}-\eqref{eq:M2-flow}. Furthermore, we fix $l^0=10^{-6}$ ($2l^0$ is the length of the dimer used to approximate the Hessian) and deem the method to have converged when $||\nabla F(\Qvec,\Mvec)||_{L^2}<10^{-6}$.

We first apply the algorithm to find an index-$k$ saddle with an appropriate initial condition, e.g. $Q_{11}=M_1=-y$ and $Q_{12}=M_2=0$, along with $n$ random orthonormal vectors as an initial eigen-frame.
Once we compute our first index-$k$ saddle $(\Qvec,\Mvec)$, the initial condition for the HiOSD method is then a perturbation of the form $(\Qvec,\Mvec)+\epsilon \mathbf{v}_i$, where $\mathbf{v}_i$ for $i=1,\ldots,n$ are the eigenvectors, and we choose $\epsilon\in(0,1]$. In principle any eigendirection is possible, but it is typically sufficient to check the first few eigendirections (starting from the eigenvector with the smallest eigenvalue). The initial eigen-frame is the set $\left\{\mathbf{v}_i\right\}$. We use this to search for an index-$\tilde{k}$ saddle. 
An index-$\tilde{k}$ saddle found using this initial condition is connected to the index-$k$ saddle point, $(\Qvec,\Mvec)$, or equivalently there is a pathway between them. Once an index-$\tilde{k}$ saddle and its eigenvectors have been found, we can use this critical point to construct new initial conditions and repeat this process to find higher, or lower index saddles and deduce the pathways between them.

\section{Numerical results}\label{sec:numerics}

We summarise the nomenclature used for the critical points of \eqref{eq:energy} in Table~\ref{tab:nomenclature}. The labels encode stability, the signs of the interior profiles of $Q_{12}$ and $M_2$, and whether a solution uses the full four-dimensional ansatz or whether a solution is of OR type (with $Q_{12} = M_2 \equiv 0$).

\begin{table}[ht]
\centering
\caption{Summary of the nomenclature used to label critical points.}
\label{tab:nomenclature}
\renewcommand{\arraystretch}{1.15}
\begin{tabularx}{\textwidth}{>{\raggedright\arraybackslash}p{0.13\textwidth}
                            >{\raggedright\arraybackslash}p{0.25\textwidth}
                            >{\raggedright\arraybackslash}X}
\toprule
Label & Meaning & Description \\
\midrule

$Stable$ 
& Stable critical point 
& A locally stable, index-$0$ critical point. \\

$+$ 
& Positive--positive profile 
& Both $Q_{12}$ and $M_2$ are positive in the interior. \\

$-$ 
& Negative--negative profile 
& Both $Q_{12}$ and $M_2$ are negative in the interior. \\

$\pm$ 
& Mixed-sign profile 
& $Q_{12}$ is positive and $M_2$ is negative in the interior. \\

$\mp$ 
& Mixed-sign profile 
& $Q_{12}$ is negative and $M_2$ is positive in the interior. \\

$F$ 
& Full solution 
& A critical point exploiting all four degrees of freedom. \\

$F i.j$ 
& Full solution index 
& Solution $j$ in full solution pair $i$; for example, $F1.2$ denotes solution $2$ in full solution pair $1$. \\

$OR$ 
& Order-reconstruction branch 
& There exists a globally stable OR branch for sufficiently large $l$. 
We denote the continuation of this solution branch as the $OR$-branch. \\

$ORi$ 
& Unstable OR branches 
& Unstable OR-type solutions that emerge as $l$ decreases. \\

\bottomrule
\end{tabularx}
\end{table}

\subsection{Effects of varying \texorpdfstring{$l$}{l}}
\label{sec:varying_l}

\begin{figure}[ht]
    \centering
    \begin{minipage}{0.38\textwidth}
        \centering
        \includegraphics[width=1.0\textwidth]{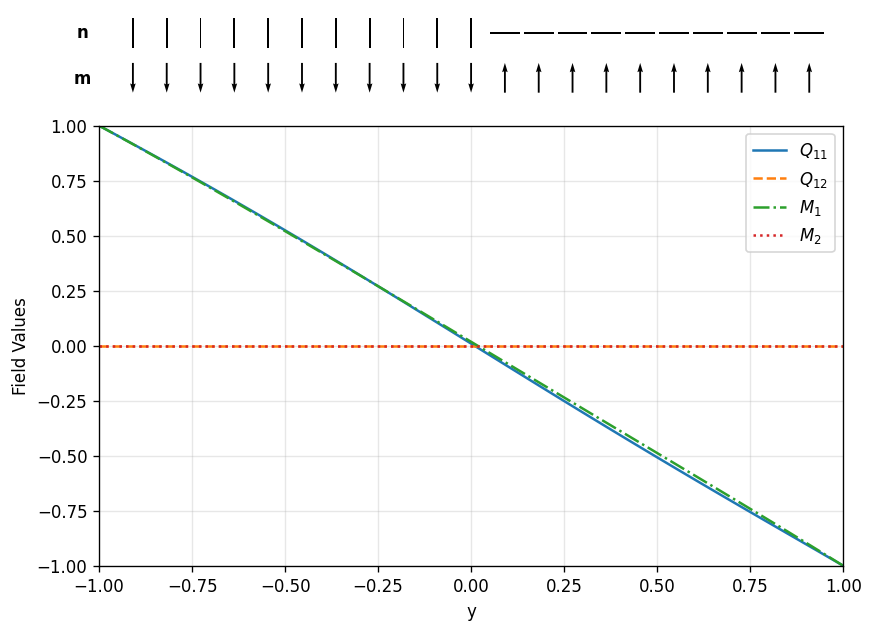}\\
        \ \textrm{$OR$ - Energy: 21.31}
    \end{minipage}
    \caption{The only critical point of \eqref{eq:energy} for $l=10$ and $c=1$. Note $Q_{11}\approx M_1\approx -y$ and $Q_{12}=M_2\equiv0$. Here we plot the components of $\Qvec$ and $\Mvec$, as well as the associated director profile $\nvec$, and normalised magnetisation vectors $\mvec=\Mvec/|\Mvec|$ (we do this for all solutions). $\nvec$ is plotted as a line field due to the physical equivalence of $\nvec$ and $-\nvec$ in the nematic phase \cite{ball_textbook}, while $\mvec$ is a unit vector and therefore plotted as arrows.} 
    \label{fig:unique_OR}
\end{figure}

We fix $c=1$ and vary $l$ to study the impact of $l$ on the solution landscape. 
From Theorem 2.5 in \cite{dalby-farrell-majumdar-xia}, for each $c>0$, there exists a positive $l^*(c)$ such that there is a unique $OR$ solution of \eqref{eq:euler-lagrange} for $l > l^*(c)$. By definition, this $OR$ solution is the unique minimiser of \eqref{eq:energy} for $l>l^*(c)$. For example, $l^*(c) = 1.25$ for $c=1$, and $l^*$ is an increasing function of $c$. 
We plot the unique $OR$ solution for $l=10$ and $c=1$ in \Cref{fig:unique_OR}, for which both $Q_{11} = M_1 \approx - y$ to match the imposed Dirichlet conditions on $y=\pm 1$. We necessarily have $\Qvec = \Mvec  = 0$ at $y \approx 0$, near the channel centre, so that the nematic and magnetic domain walls coincide at the channel centre. We note that $\nvec \approx (\pm 1, 0)$ for $y \leq 0$ and $\nvec \approx (0, \pm 1)$ for $y >0$. Similarly, $\mvec = (1, 0)$ for $y \leq 0$ and $\mvec = (-1, 0)$ for $y>0$. From Theorem 3.3 in \cite{dalby-farrell-majumdar-xia}, this $OR$ solution loses stability as $l$ decreases, for a fixed $c>0$. 
Stable full solutions therefore emerge when this $OR$ solution loses stability, and hence we focus on $l \leq 1$ in this paper.

\subsubsection{\texorpdfstring{$l=1$}{l=1} and \texorpdfstring{$c=1$}{c=1}}

\begin{figure}[ht]
    \centering
    \begin{minipage}{0.38\textwidth}
        \centering
        \includegraphics[width=1.0\textwidth]{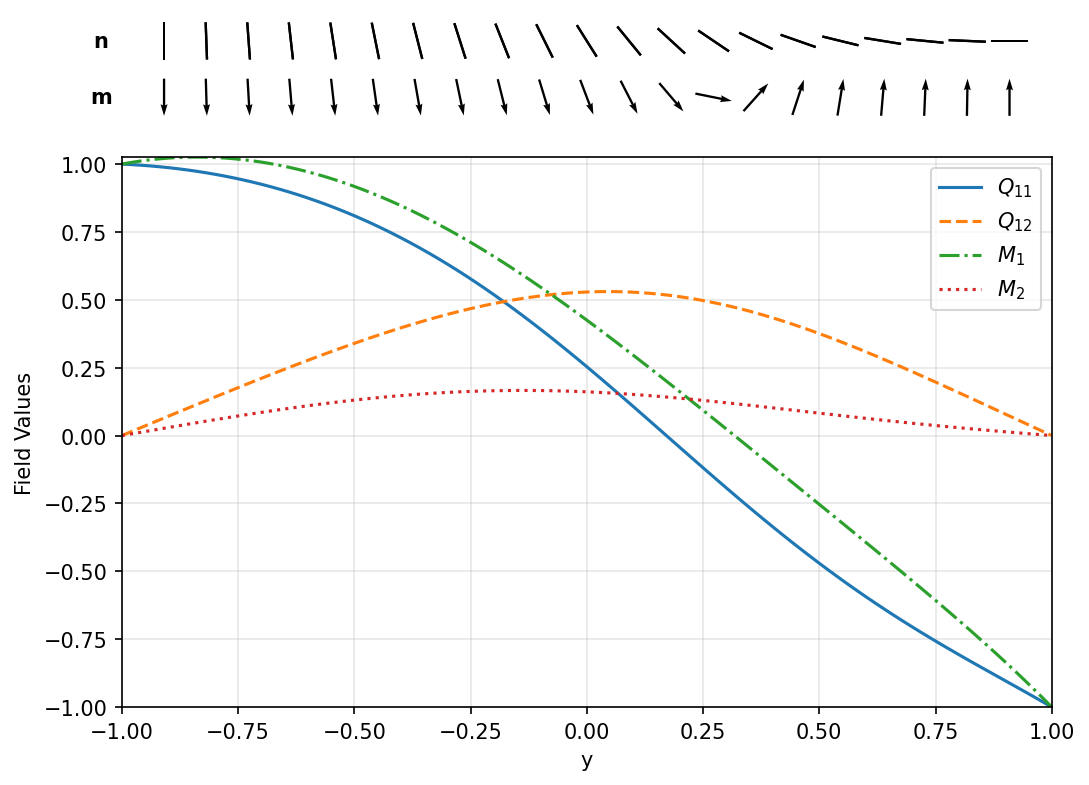}\\
        \ \textrm{$Stable+$}
    \end{minipage}
    \begin{minipage}{0.38\textwidth}
        \centering
        \includegraphics[width=1.0\textwidth]{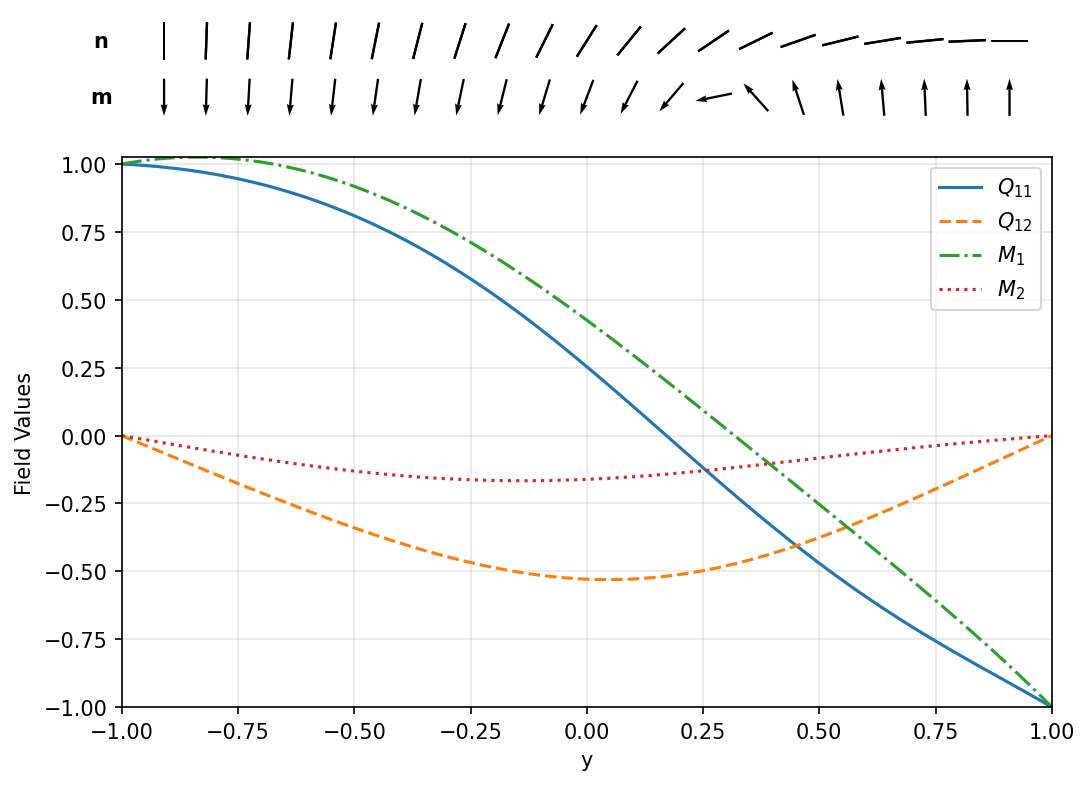}\\
        \ \textrm{$Stable-$}
    \end{minipage}
    \begin{minipage}{0.38\textwidth}
        \centering
        \includegraphics[width=1.0\textwidth]{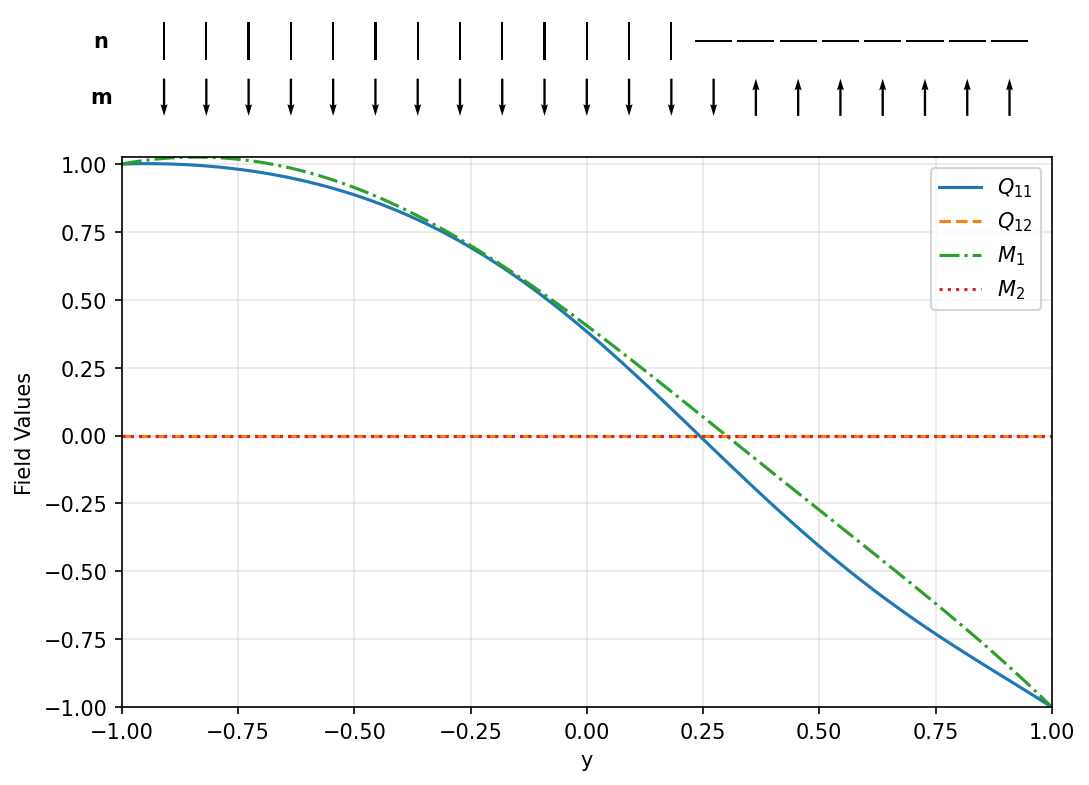}\\
        \ \textrm{$OR$ - index-1 saddle}
    \end{minipage}
    \begin{minipage}{0.38\textwidth}
        \centering
        \includegraphics[width=1\textwidth]{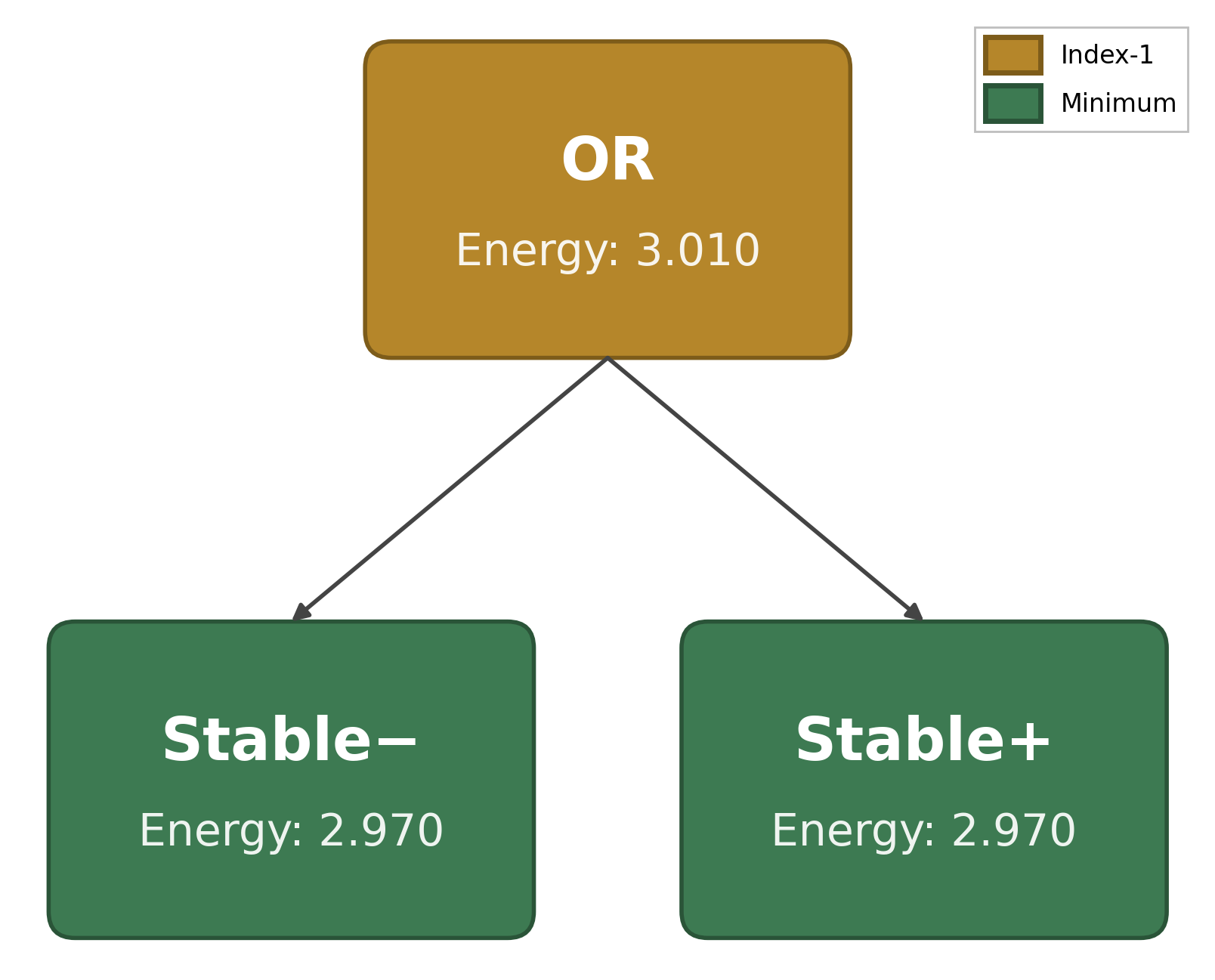}
    \end{minipage}
    \caption{The critical points of \eqref{eq:energy} for $l=1$ and $c=1$ and the associated solution landscape.} 
    \label{fig:saddles_l=1_c=1}
\end{figure}

For $l=c=1$, the $OR$ solution exists but becomes unstable, accompanied by the emergence of two stable and energetically degenerate full solutions that exploit the four degrees of freedom, as plotted in \Cref{fig:saddles_l=1_c=1}.
For clarity, $Stable+$ and $Stable-$ belong to a full solution pair in \Cref{fig:saddles_l=1_c=1}; $Stable+$ and $Stable-$ have the same $Q_{11}$ and $M_1$ profiles, but the associated $Q_{12}$ and $M_2$ profiles are reflections of each other in the $y$-axis. Let $\theta$ denote the director angle for $Stable+$ and let $\psi$ denote the director angle for $Stable-$, then $\theta+\psi=0$ on $\Omega$. Hence, the director is $\nvec=(n_1,n_2)$ for $Stable+$, and $\nvec=(n_1,-n_2)$ for $Stable-$. 
The corresponding normalised magnetisation profiles $\mvec = \frac{\Mvec}{|\Mvec|}$ also vary in their sense of rotation, since the $M_2$ profiles have opposite signs. 
This is true for subsequent full solution pairs in this paper.

We find one index-1 $OR$ solution (plotted in \Cref{fig:saddles_l=1_c=1}), which is the continuation of the unique $OR$ solution for large $l$. 
The connectivity is simple in this case: $Stable+$ and $Stable-$ are connected via the index -$1$ $OR$ solution (see \Cref{fig:saddles_l=1_c=1}).  Consider the pathway from $Stable+$ to $Stable-$; $Q_{12}$ and $M_2$ collapse to the zero functions everywhere on the $OR$ solution and we see nematic and magnetic polydomains in the $OR$ solution. This allows for $Q_{12}$ and $M_2$ to reverse sign along the pathway connecting the $OR$ solution to the final $Stable-$ solution. This example highlights the importance of OR solutions in switching processes since the $OR$ solution is the only available transition state for this example. 

\subsubsection{\texorpdfstring{$l=0.2$}{l=0.2} and \texorpdfstring{$c=1$}{c=1}}\label{sec:l=0.2_c=1}
We take $l=0.2$ and $c=1$ in this section to further investigate the effects of decreasing $l$ on the ferronematic solution landscapes.

\begin{figure}[ht]
    \centering
    \begin{minipage}{0.38\textwidth}
        \centering
        \includegraphics[width=1.0\textwidth]{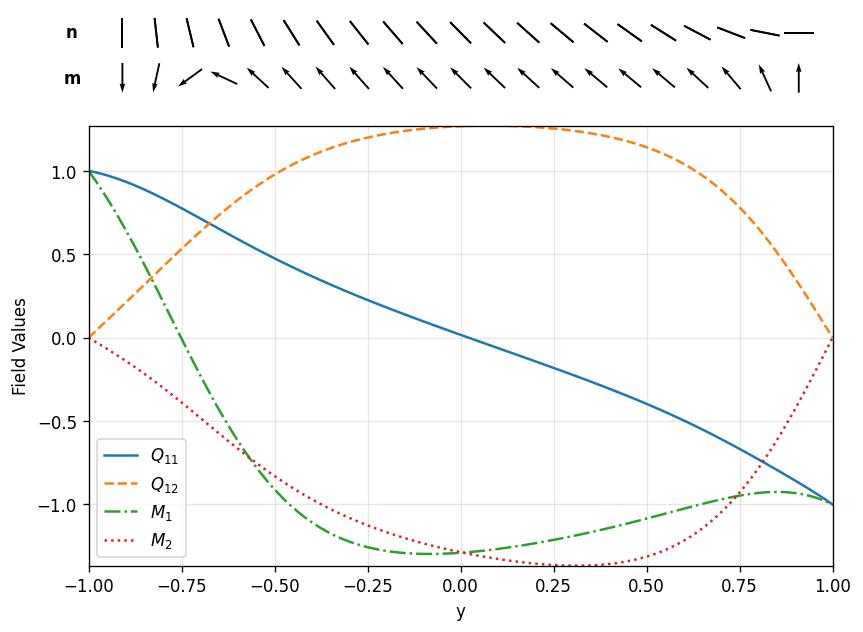}\\
        \ \textrm{$Stable\pm$}
    \end{minipage}
    \begin{minipage}{0.38\textwidth}
        \centering
        \includegraphics[width=1.0\textwidth]{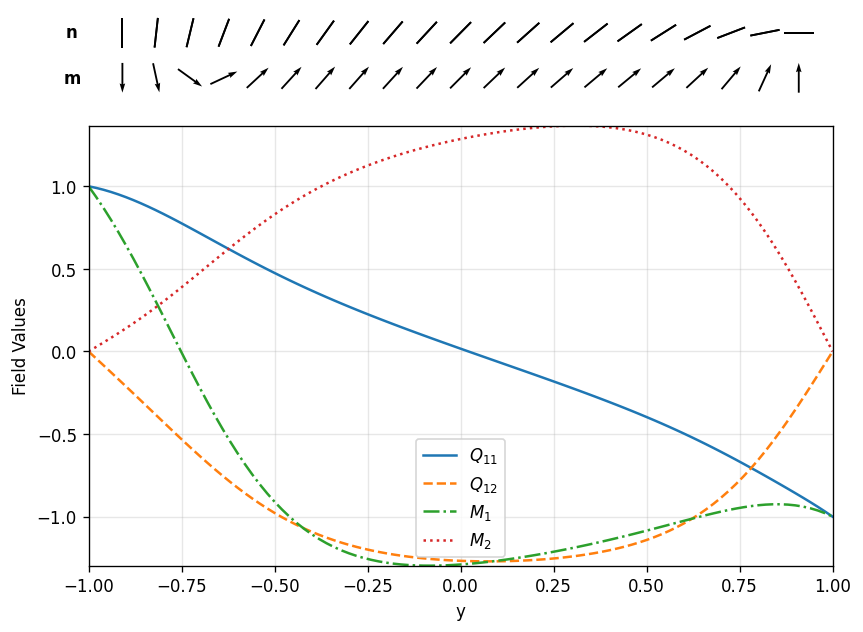}\\
        \ \textrm{$Stable\mp$}
    \end{minipage}
    \begin{minipage}{0.38\textwidth}
        \centering
        \includegraphics[width=1.0\textwidth]{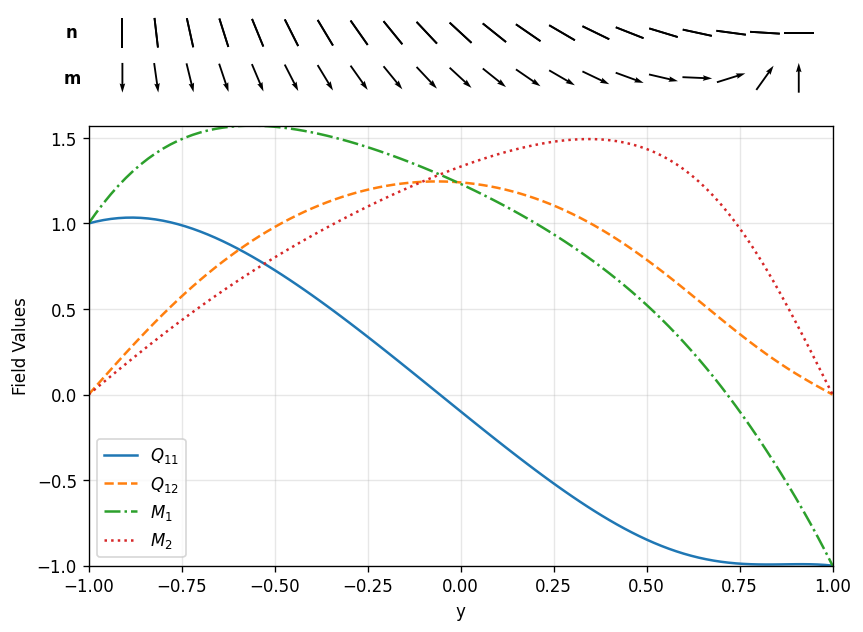}\\
        \ \textrm{$Stable+$}
    \end{minipage}
    \begin{minipage}{0.38\textwidth}
        \centering
        \includegraphics[width=1.0\textwidth]{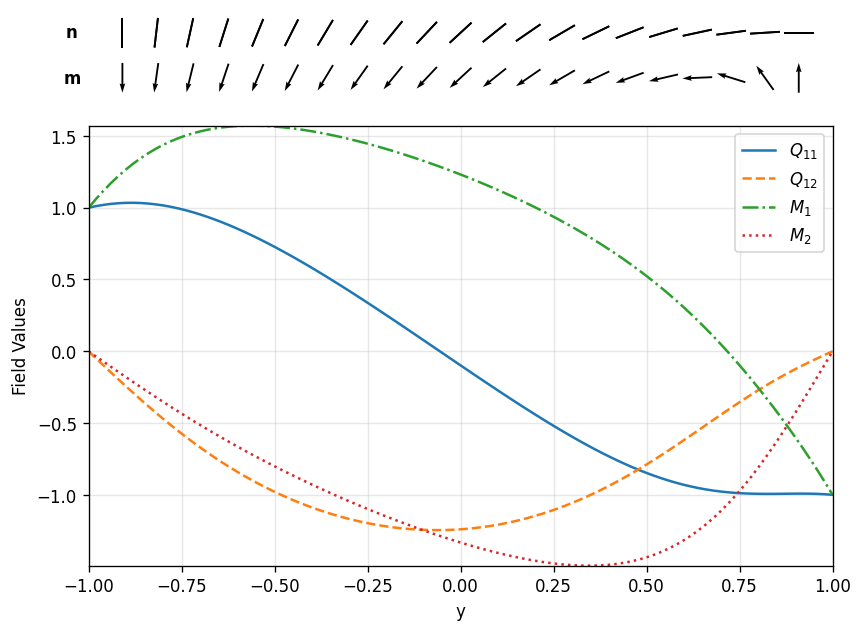}\\
        \ \textrm{$Stable-$}
    \end{minipage}
    \caption{The stable critical points of \eqref{eq:energy} for $l=0.2$ and $c=1$. $Stable+$ and $Stable-$ have equal and lowest energy, while $Stable\pm$ and $Stable\mp$ also have equal energy (see \Cref{fig:flow_diagram_l=0.2_c=1} for energy values).}
    \label{fig:stable_l=0.2_c=1}
\end{figure}
For $l=0.2$ and $c=1$, there are two stable solution pairs of the ferronematic free energy \eqref{eq:energy}, as in \Cref{fig:stable_l=0.2_c=1}.  
The two pairs are $(Stable +, Stable -)$ and $(Stable \pm, Stable \mp)$. $(Stable +, Stable -)$ are energetically degenerate and have the lowest energy; both exhibit a continuous and smooth rotation of $\nvec$ between $y=\pm 1$. $(Stable \pm, Stable \mp)$
has higher energy than $(Stable+, Stable -)$ because there are sharp rotation gradients or elastic penalties in $\mvec$ near one of the edges.

\begin{figure}[ht!]
    \centering
    \begin{minipage}{0.38\textwidth}
        \centering
        \includegraphics[width=1.0\textwidth]{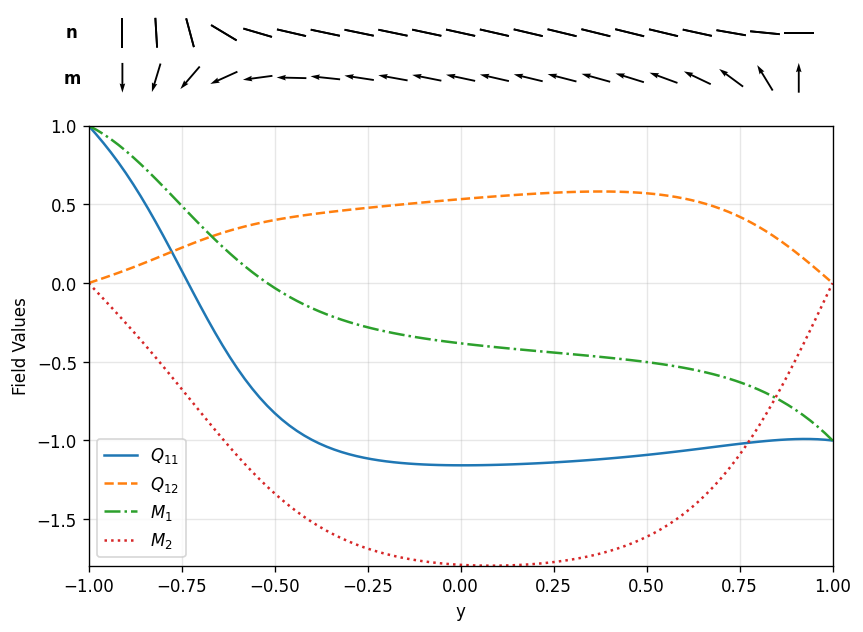}\\
        \ \textrm{$F1.1$ - index-1 saddle}
    \end{minipage}
    \begin{minipage}{0.38\textwidth}
        \centering
        \includegraphics[width=1.0\textwidth]{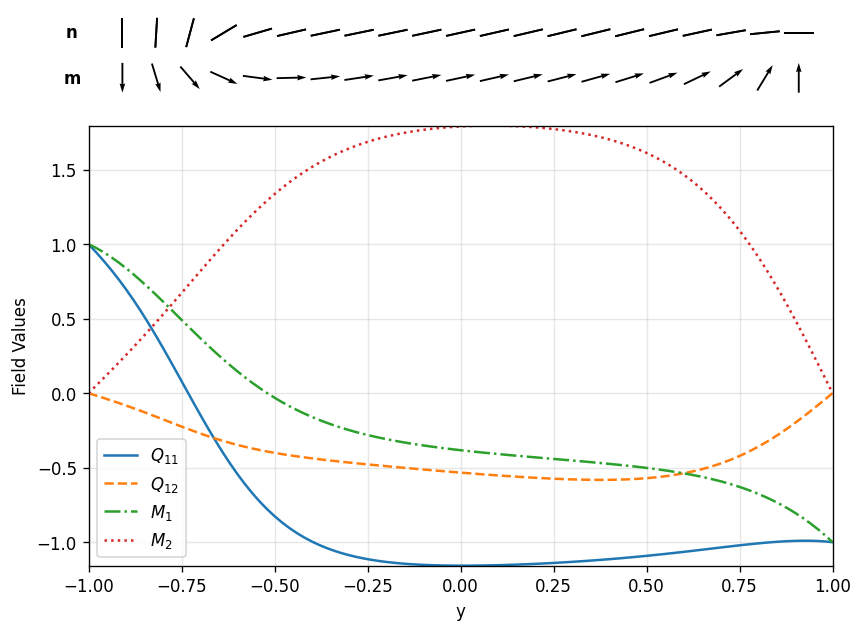}\\
        \ \textrm{$F1.2$ - index-1 saddle}
    \end{minipage}
    \begin{minipage}{0.38\textwidth}
        \centering
        \includegraphics[width=1.0\textwidth]{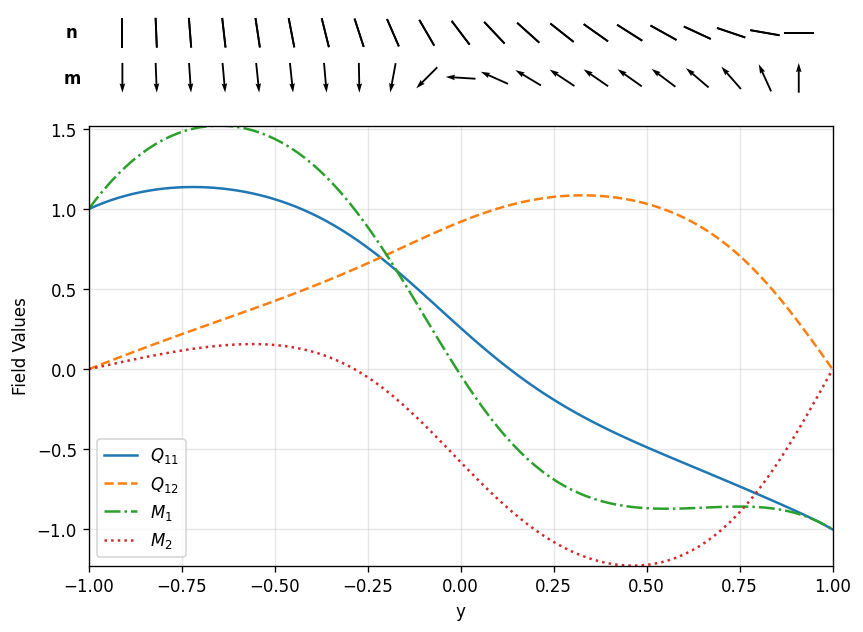}\\
        \ \textrm{$F2.1$ - index-1 saddle}
    \end{minipage}
    \begin{minipage}{0.38\textwidth}
        \centering
        \includegraphics[width=1.0\textwidth]{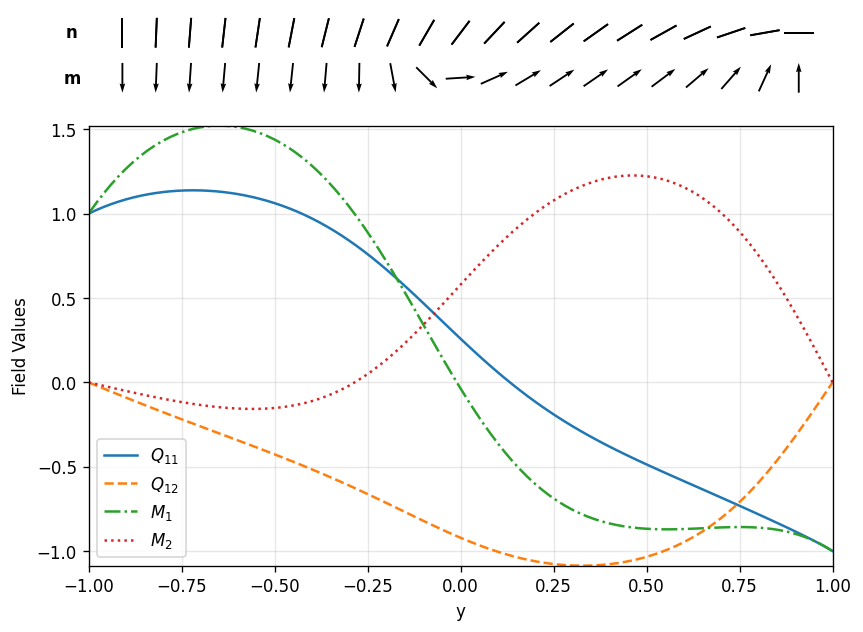}\\
        \ \textrm{$F2.2$ - index-1 saddle}
    \end{minipage}
    \begin{minipage}{0.38\textwidth}
        \centering
        \includegraphics[width=1.0\textwidth]{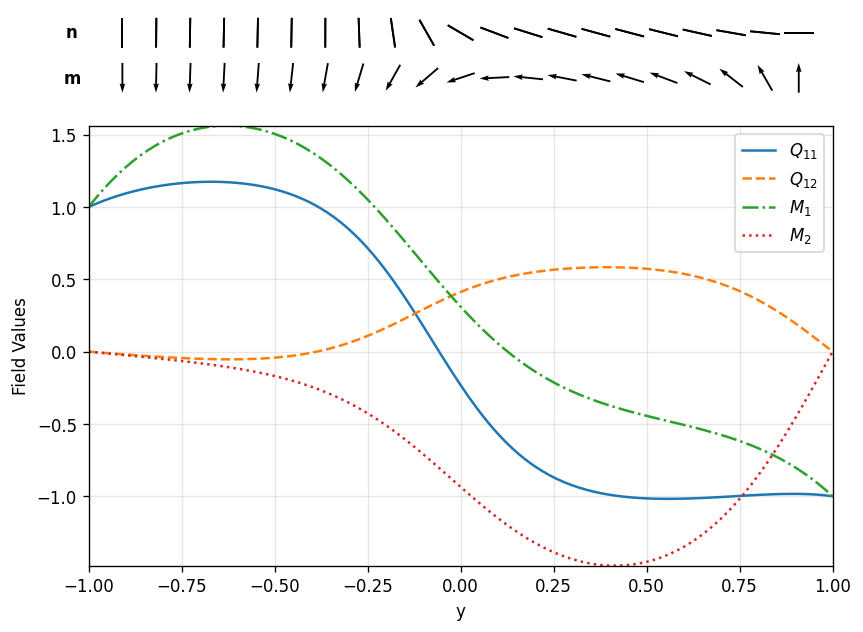}\\
        \ \textrm{$F3.1$ - index-2 saddle}
    \end{minipage}
    \begin{minipage}{0.38\textwidth}
        \centering
        \includegraphics[width=1.0\textwidth]{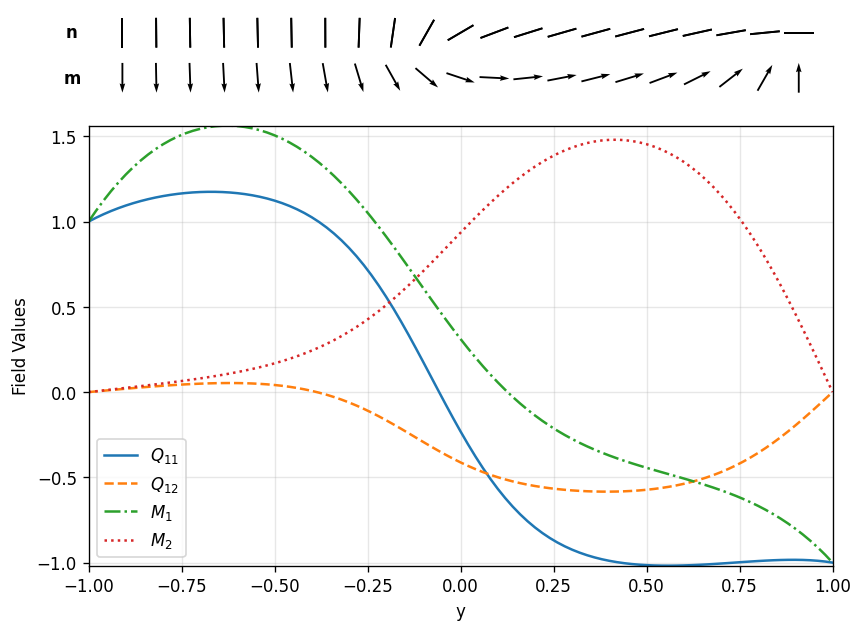}\\
        \ \textrm{$F3.2$ - index-2 saddle}
    \end{minipage}
    \caption{Unstable full saddle points of the energy \eqref{eq:energy} for $l=0.2$ and $c=1$. The order of the energy of these saddle points is $F1<F2<F3$ (see \Cref{fig:flow_diagram_l=0.2_c=1} for energy values).}
    \label{fig:full_saddle_points_l=0.2_c=1}
\end{figure}

\begin{figure}[ht!]
    \centering
    \begin{minipage}{0.4\textwidth}
        \centering
        \includegraphics[width=1.0\textwidth]{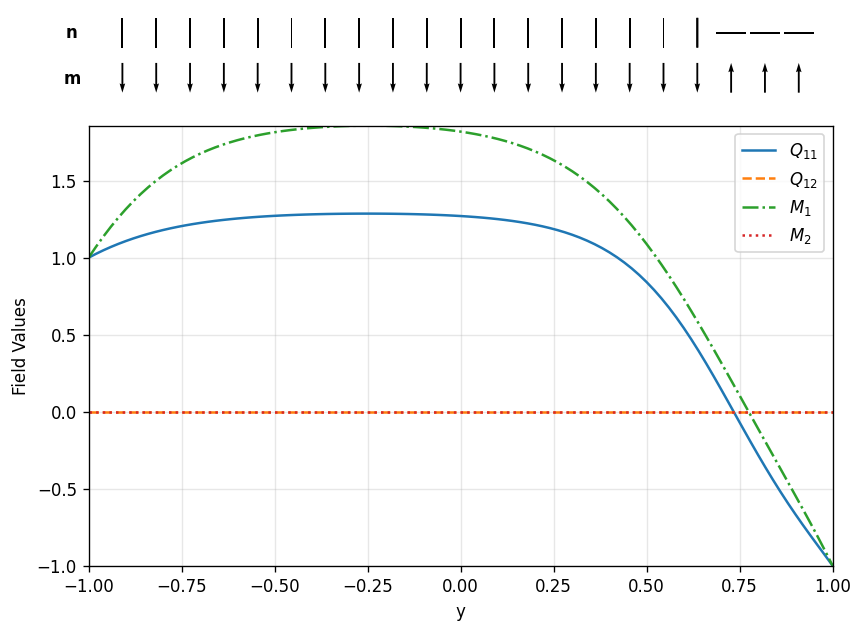}\\
        \ \textrm{$OR$  - index-1 saddle}
    \end{minipage}
    \begin{minipage}{0.4\textwidth}
        \centering
        \includegraphics[width=1.0\textwidth]{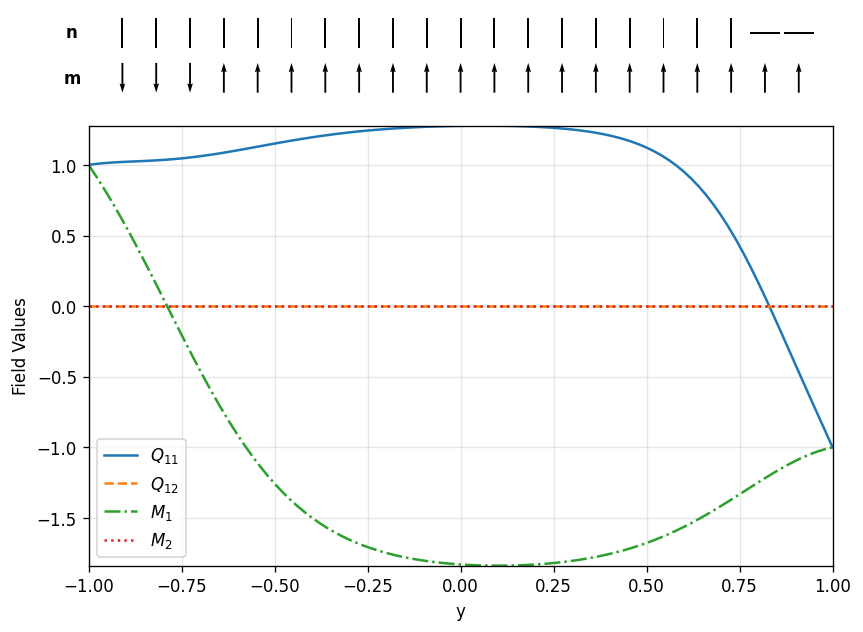}\\
        \ \textrm{$OR1$  - index-1 saddle}
    \end{minipage}
    \begin{minipage}{0.4\textwidth}
        \centering
        \includegraphics[width=1.0\textwidth]{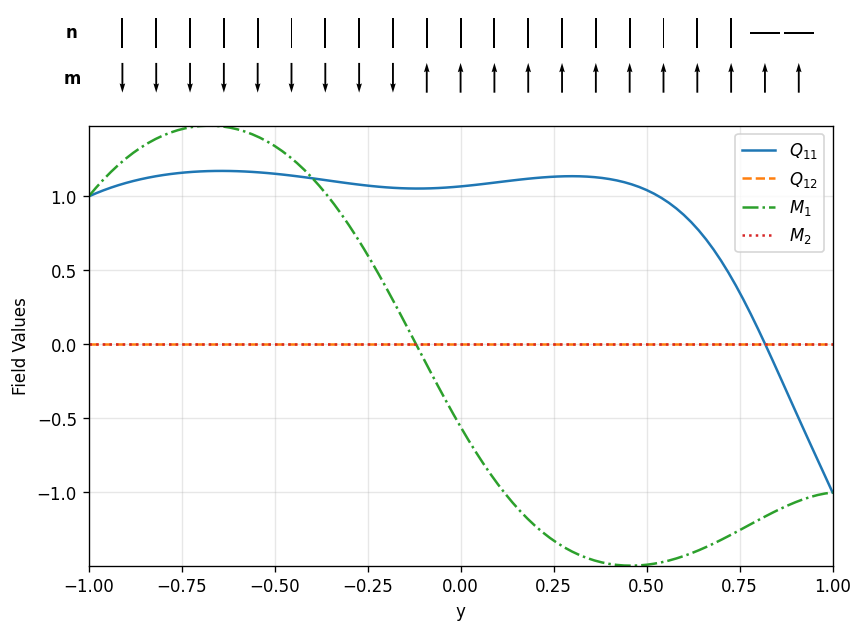}\\
        \ \textrm{$OR2$  - index-2 saddle}
     \end{minipage}
     \caption{Unstable OR saddle points of the energy \eqref{eq:energy} for $l=0.2$ and $c=1$. The energies are ordered as $OR<OR1<OR2$ (see \Cref{fig:flow_diagram_l=0.2_c=1} for energy values).}
    \label{fig:OR_saddle_points_l=0.2_c=1}
\end{figure}

There are nine saddle points of the free energy \eqref{eq:energy} as plotted in \Cref{fig:full_saddle_points_l=0.2_c=1} and \Cref{fig:OR_saddle_points_l=0.2_c=1}, along with their indices. There are three full solution pairs of saddle points and three OR saddle points of \eqref{eq:energy}. The Morse index of full solutions is the same within each solution pair. The saddle points can be ordered in terms of their energies as $OR<F1<OR1<F2<F3<OR2$, where $OR$ is the continuation of the unique $OR$ solution branch for $l=10, c=1$.

The full solution pairs are plotted in \Cref{fig:full_saddle_points_l=0.2_c=1}: $F1.1$ and $F1.2$ (both index-1), $F2.1$ and $F2.2$ (both index-1), and $F3.1$ and $F3.2$ (both index-2). Looking at the director and magnetisation plots for the saddle points, the rotations (needed to match the conflicting boundary conditions for $\nvec$ and $\Mvec$) are localised near the edges for $F1$; the $\mvec$ profiles exhibit a sharp rotation gradient near the channel centre for $F2$ but the $\nvec$ plots do not exhibit any sharp gradients as such. With regards to the higher-index $F3$ pair, both $\nvec$ and $\mvec$ exhibit a localised jump near the channel centre which might explain the higher Morse index. 

There are also three OR solutions shown in \Cref{fig:OR_saddle_points_l=0.2_c=1}: $OR$ and $OR1$, which are both index-1 saddles, and $OR2$ which is an index-2 saddle point. $OR$ is the continuation of the unique OR solution branch for large $l$, while $OR1$ and $OR2$ are new critical points. We notice that all three OR solutions have one nematic and magnetic domain wall, which need not coincide. Notably, the nematic and magnetic domain walls are close to each other, near $y=1$ for the $OR$ critical point. In the case of $OR1$, the nematic domain wall (with $\Qvec=0$) is located near $y=1$ whilst the magnetic domain wall (with $M_1 = M_2 = 0$) is located near $y=-1$. In the case of the higher-index $OR2$ critical point, the magnetic domain wall is near $y=0$ and the nematic domain wall is located near $y=1$. We speculate that the Morse index of a critical point increases when the domain walls are located near the channel centre. 
In \cite{dalby-farrell-majumdar-xia}, the authors do some asymptotic analysis of OR solutions for small $l$.
They show that physically relevant OR solutions (which are candidates for either stable or index-$1$ critical points of \eqref{eq:energy}) converge to $(Q_{11},M_1)=(\rho^*,\pm\sqrt{1+2c\rho^*})$ (recall that $Q_{12}=M_2=0$ everywhere for OR solutions) almost everywhere,  in the $l\to 0$ limit, where 
\begin{multline}
    \rho^*=\left(\frac{c}{8}+\sqrt{\frac{c^2}{64}-\frac{1}{27}\left(1+\frac{c^2}{2}\right)^3}\right)^\frac{1}{3}\\
    +\left(\frac{c}{8}-\sqrt{\frac{c^2}{64}-\frac{1}{27}\left(1+\frac{c^2}{2}\right)^3}\right)^\frac{1}{3}.
\end{multline}
For $c=1$, $\rho^*=1.3008$ and $\sqrt{1+2c\rho^*}=1.8978$. 
 It follows that $OR$ is approximately equal to $\pvec^*:=(\rho^*,\sqrt{1+2c\rho^*})$ in the interior, $OR1$ is approximately equal to $\pvec^{**}:=(\rho^*,-\sqrt{1+2c\rho^*})$ in the interior, whilst $OR2$ jumps from $\pvec^*$ to $\pvec^{**}$, via a transition layer. So, it is likely that $OR2$ has higher index because it has an internal transition layer (and central magnetic domain wall due to $M_1=0$ at $y\approx-0.17$) which is energetically expensive.

\begin{figure}
    \centering
    \includegraphics[width=1.0\textwidth]{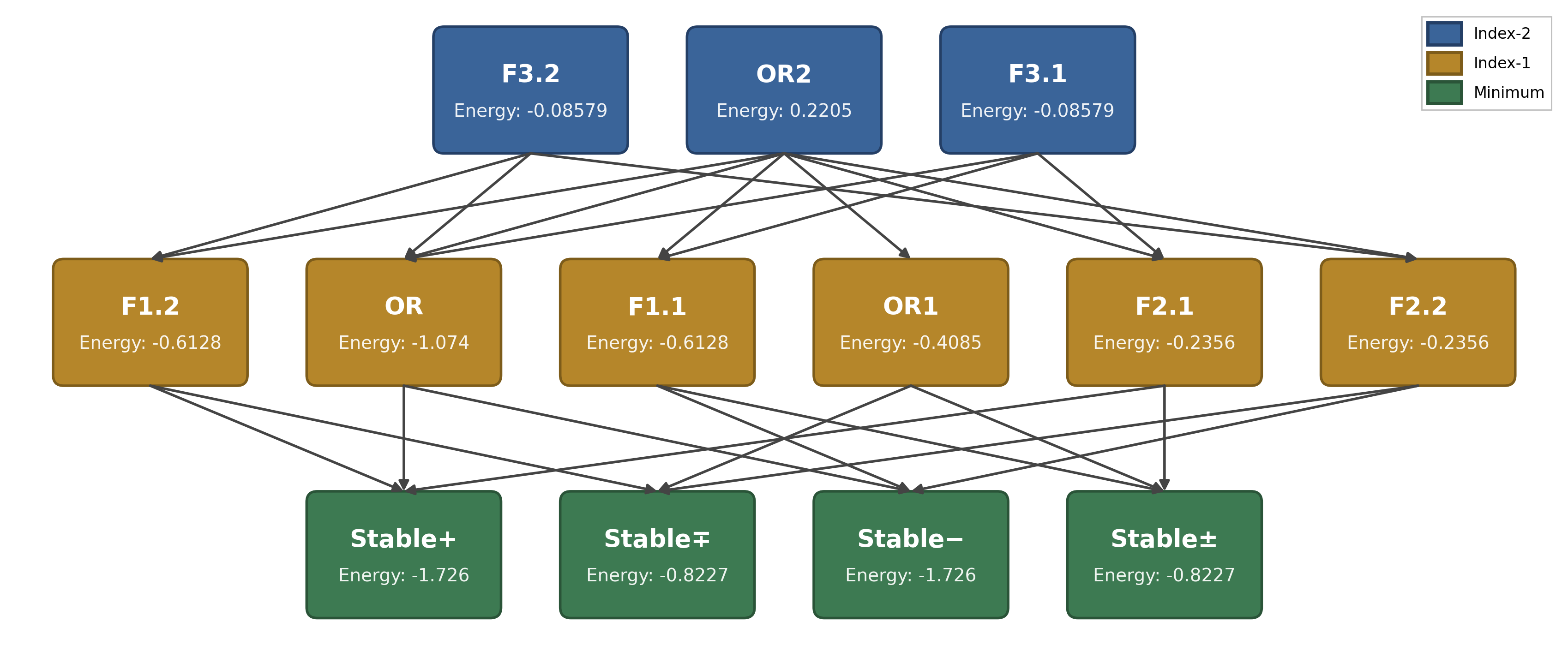}
    \caption{Solution landscape for $l=0.2$ and $c=1$. We only include arrows showing how an index-$i$ saddle point connects to any index-$i+1$ saddle point. That is not to say there do not exist pathways between index-$i$ and index-$i+2$ saddle points. We do not include arrows for these pathways since our diagram becomes cluttered and difficult to interpret. 
    }
    \label{fig:flow_diagram_l=0.2_c=1}
\end{figure}

In \Cref{fig:flow_diagram_l=0.2_c=1}, we study the connectivity of the solution landscape, i.e., how the four stable solutions in \Cref{fig:stable_l=0.2_c=1} connect to the unstable saddle points in \Cref{fig:full_saddle_points_l=0.2_c=1} and \Cref{fig:OR_saddle_points_l=0.2_c=1}. There are some important points: $OR2$ is the parent state (the state with highest index, which connects to all solutions with lower Morse index, i.e., $OR2$ connects to all index-1 states in \Cref{fig:flow_diagram_l=0.2_c=1}). 
We can also use $OR2$ to connect any pair of stable solutions in \Cref{fig:flow_diagram_l=0.2_c=1}. The $OR$ solution is the lowest energy index-$1$ saddle point that connects the global energy minimisers, $Stable+$ and $Stable-$. The index-$1$ $OR1$ critical point connects the stable full solution pair, $Stable\pm$ to $Stable\mp$. We speculate that the $OR$-mediated pathway between $Stable+$ and $Stable-$ is most likely to be observed in switching processes. The $F1$ and $F2$ are also index-$1$ saddle points that connect full stable solutions in different families, i.e., $Stable+$ to $Stable \mp$, etc. Thus, the OR saddle points and the $F1, F2$ saddle points serve different purposes for switching processes, and this warrants further study in the future.

\subsubsection{\texorpdfstring{$l=0.1$}{l=0.1} and \texorpdfstring{$c=1$}{c=1}}\label{sec:l=0.1_c=1}
Next, we consider $l=0.1$ and $c=1$ and find a total of thirty-nine critical points of \eqref{eq:energy}: $2$ pairs of full stable solutions, $15$ pairs of unstable full solutions and $5$ unstable OR solutions. A pair of index-$5$ full solutions are the highest index solutions.

\begin{figure}
    \centering
    \begin{minipage}{0.38\textwidth}
        \centering
        \includegraphics[width=1.0\textwidth]{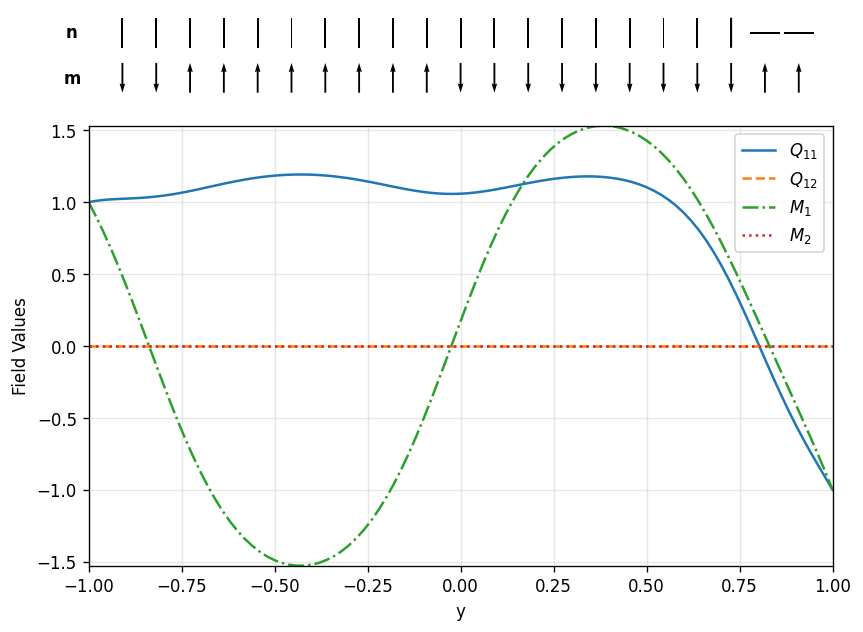}\\
        \ \textrm{$OR3$ - index-2 saddle}
    \end{minipage}
    \begin{minipage}{0.38\textwidth}
        \centering
        \includegraphics[width=1.0\textwidth]{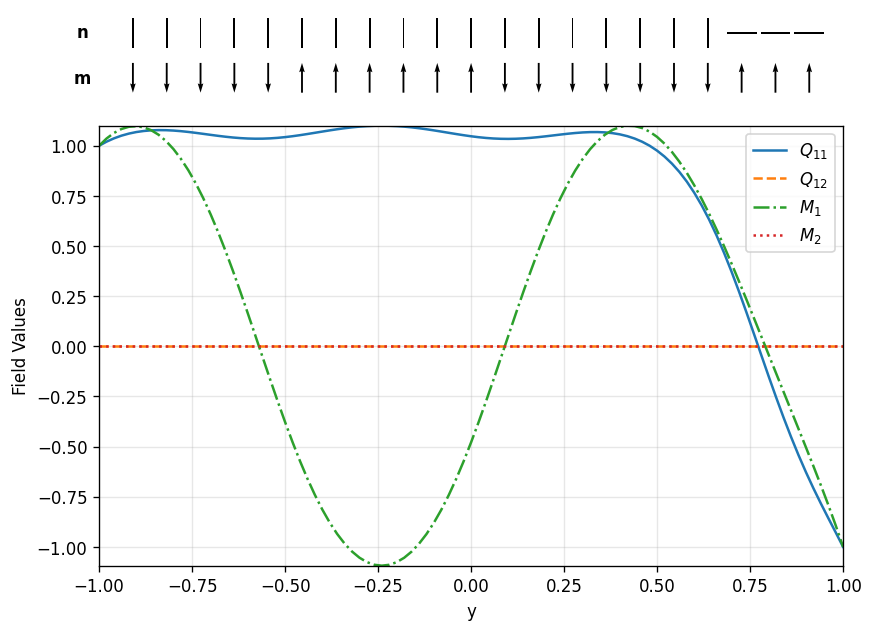}\\
        \ \textrm{$OR4$ - index-3 saddle}
    \end{minipage}
    \caption{Two unstable OR saddle points of the energy \eqref{eq:energy} for $l=0.1$ and $c=1$. $OR3$ has energy -1.21, and $OR4$ has energy 1.083.}
    \label{fig:OR_saddles_l=0.1_c=1}
\end{figure}

We plot two new high-index OR solutions in \Cref{fig:OR_saddles_l=0.1_c=1}: $OR3$ (index-$2$) and $OR4$ (index-$3$), both of which have three magnetic domain walls. The $OR3$ and $OR4$ profiles differ in the location of magnetic domain walls, such that two domain walls are shifted towards the channel interior/centre for the $OR4$ solution. This again supports the idea that a higher Morse index is associated with a larger number of interior magnetic and/or nematic domain walls. We omit the solution landscape plot for brevity. 

The key point is that there is a sharp increase in the number of critical points of \eqref{eq:energy} as $l$ decreases, including full solutions and OR solutions. The total number of numerically computed critical points of \eqref{eq:energy} increases from $1\to 3\to 13 \to 39$, as we decrease $l$ from $10\to 1 \to0.2\to 0.1$, for a fixed $c=1$.  For OR solutions, the multiplicity goes from $1\to1\to3\to5$ as $l$ decreases from $10\to 1 \to0.2\to 0.1$. The multiplicity of full solutions increases from $0\to2\to10\to34$, as $l$ decreases from $10\to 1 \to0.2\to 0.1$. The rescaled parameter, $l$ is inversely proportional to the channel width $D^2$, so that wider channels support a larger class of index-$1$ and higher-index OR solutions, that are potentially relevant for non-equilibrium and transport processes. 

\subsection{Effects of varying \texorpdfstring{$c$}{c}}\label{sec:varying_c}
In \cite{dalby-farrell-majumdar-xia}, the authors speculate that increasing $c$ (the strength of the nemato-magnetic coupling) destabilises OR solutions. Therefore, we hypothesise that increasing $c$ has the same effect as decreasing $l$ on the solution landscapes, e.g., the solution landscapes may be qualitatively similar for two given pairs, $(\hat{l}, \hat{c})$ and $(\tilde{l}, \tilde{c})$ if the ratios $\frac{\hat{l}}{\hat{c}}$ and $\frac{\tilde{l}}{\tilde{c}}$ are comparable. 
To test this, we recall some basic facts about the $c=0$ landscape, and then vary $c$ for fixed $l=1$. 

\subsubsection{c=0}
We perform a few numerical experiments with $c=0$, $l=\left\{10,1, 0.2 \right\}$ to complement the work in the previous section with $c=1$. The solution landscape is unchanged for $l=10$ and $l=1$, i.e., we recover the unique $OR$ solution with $(l,c)=(10,0)$, and two full stable solutions ($Stable+$ and $Stable-$) connected by an index-$1$ $OR$ solution for $(l,c)=(1, 0)$.

\begin{figure} 
\centering
\includegraphics[width=0.75\textwidth]      {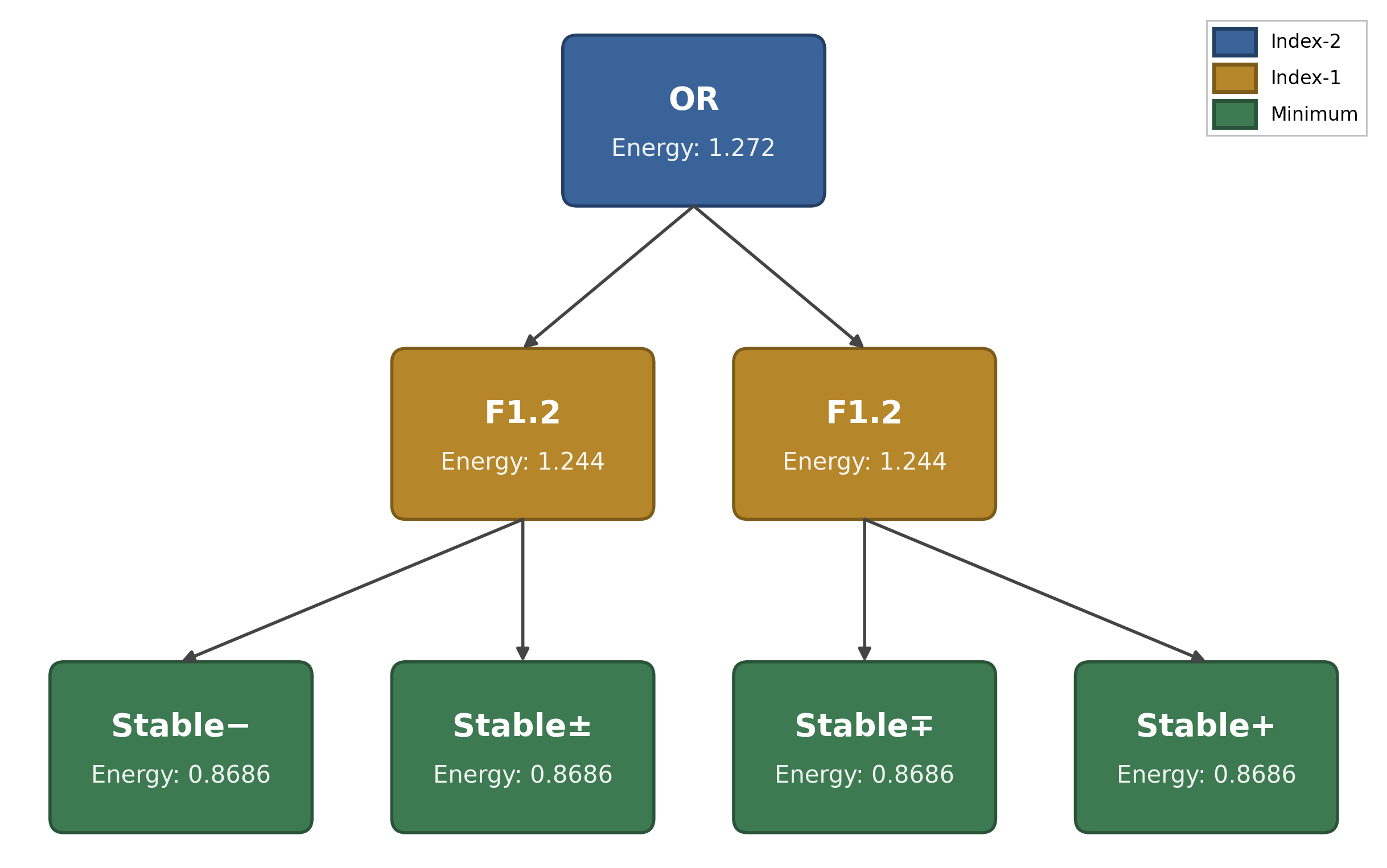}
    \caption{Solution landscape for $l=0.2$ and $c=0$.}
    \label{fig:flow_diagrams_c=0}
\end{figure}

There are differences between the solution landscapes for $(l,c)=(0.2,0)$ and $(l,c) = (0.2,1)$. Namely, for $c=0$, there are the four familiar stable solutions, connected by a pair of index-$1$ full solutions, which are further connected by the parent index-$2$ $OR$ solution; see \Cref{fig:flow_diagrams_c=0}. By contrast, there are $6$ unstable full solutions and $3$ unstable OR solutions for $(l,c) = (0.2,1)$. Next, we present some heuristic arguments to explain the increase in the multiplicity of critical points of \eqref{eq:energy} with non-zero $c$.

When $c = 0$, the free energy \eqref{eq:energy} decouples as $F = F_{\mathbf{Q}}[\Qvec] + F_{\mathbf{M}}[\Mvec]$, and the two fields are independent.
Define the pointwise transformations
\begin{align}
  &\sigma_Q \colon (Q_{11}, Q_{12}, M_1, M_2) \mapsto (Q_{11}, -Q_{12}, M_1, M_2),
  \\
  &\sigma_M \colon (Q_{11}, Q_{12}, M_1, M_2) \mapsto (Q_{11}, Q_{12}, M_1, -M_2),
  \label{eq:alpha_beta}
\end{align}
and set $\sigma := \sigma_Q \circ \sigma_M$, so that $\sigma$ negates both
$Q_{12}$ and $M_2$ simultaneously. All three transformations preserve the imposed Dirichlet conditions for $\Qvec$ and $\Mvec$ on $y=\pm 1$ (see Table~\ref{table:BCs-ferronematics}).
 
Recall that the \emph{orbit} of a critical point $\mathbf{u}^*$ of \eqref{eq:energy} under a
group $G$ is the set $G \cdot \mathbf{u}^* = \{g \cdot \mathbf{u}^* : g \in
G\}$, and that every element of this orbit is also a critical point with the
same energy, since the free energy is $G$-invariant. Two critical points in the same
orbit are related by a symmetry of the system and represent the same
physical configuration; the number of \emph{distinct} physical configurations is
therefore the number of orbits, not the total count of critical points of \eqref{eq:energy}.  
Take any full solution $\mathbf{u}^* = (Q_{11}^*, Q_{12}^*, M_1^*, M_2^*)$
of the $c = 0$ system with $Q_{12}^* \not\equiv 0$ and $M_2^* \not\equiv
0$. Its $G_0$-orbit has size four:
\[
  \{ \mathbf{u}^*,\; \sigma_Q(\mathbf{u}^*),\; \sigma_M(\mathbf{u}^*),\;
     \sigma(\mathbf{u}^*) \}.
\]
These four functions are distinct but form a single equivalence class under
the symmetry group of the $c = 0$ system. However, one can use group-theoretic arguments to show that the single $G_0$-orbit of size four at $c = 0$ becomes two
$G_c$-orbits of size two at $c \neq 0$: the number of distinct equivalence classes (under the system's own symmetry group) doubles, partially explaining the increase of critical points of \eqref{eq:energy} for $c \neq 0$. These arguments only apply to full solutions of \eqref{eq:euler-lagrange} that are non-degenerate at $c = 0$.
Non-degeneracy is a generic condition and can be verified numerically by checking that the Hessian matrix of \eqref{eq:energy} has no zero eigenvalue. 
An analogous orbit-splitting mechanism for coupled nonlinear Schr\"{o}dinger systems, which share the same product-symmetry structure, is discussed in \cite{Dancer2010}. 
 
A second, independent mechanism also contributes to the increase in critical
points of \eqref{eq:energy} for $c\neq 0$: the coupling modifies the Hessian of \eqref{eq:energy} along an OR solution branch, and as
$c$ increases, an eigenvalue can cross zero, producing new critical points
with $Q_{12},M_2 \not\equiv 0$ via a pitchfork bifurcation from an OR solution branch.
These are genuinely new critical points with no $c = 0$ ancestor. 

 We can also get a complementary viewpoint by means of separability and angular momentum exchange-based arguments. For $c = 0$, the energy \eqref{eq:energy} can be decomposed as $F_0(\mathbf{Q}, \mathbf{M}) =
F_{\mathbf{Q}}(\mathbf{Q}) + F_{\mathbf{M}}(\mathbf{M})$, so that a critical
point must satisfy two \emph{independent} decoupled Euler--Lagrange systems,
\begin{equation}
  R_{\mathbf{Q}}(\mathbf{Q}) = 0, \qquad R_{\mathbf{M}}(\mathbf{M}) = 0,
  \label{eq:decoupled_EL}
\end{equation}
where $R_{\mathbf{Q}}$ and $R_{\mathbf{M}}$ are the Euler--Lagrange
residuals of $F_{\mathbf{Q}}$ and $F_{\mathbf{M}}$ respectively.  Every
solution of the full system at $c = 0$ is therefore a \emph{product} of a
solution of the $\mathbf{Q}$-subsystem and a solution of the
$\mathbf{M}$-subsystem. For $c \neq 0$, equations \eqref{eq:Q11_again}--\eqref{eq:M2_again} show that the residuals become
\begin{equation}
  R_{\mathbf{Q}}(\mathbf{Q}) = c\,\mathbf{a}(\mathbf{M}), \qquad
  R_{\mathbf{M}}(\mathbf{M}) = c\,\mathbf{b}(\mathbf{Q}, \mathbf{M}),
  \label{eq:coupled_EL}
\end{equation}
where $\mathbf{a}(\mathbf{M}) = \bigl(M_1^2 - M_2^2,\; 2M_1 M_2\bigr)$ and
$\mathbf{b}(\mathbf{Q},\mathbf{M})$ collects the $c$-terms from \eqref{eq:M1_again}--\eqref{eq:M2_again}.  The
$c$-terms are now \emph{source terms} rather than additional constraints. 
The solution set is no longer
a Cartesian product of two independent boundary-value problems, and new solutions are possible by means of mutual balancing. 
 
To see the geometric content, we can write $\mathbf{Q}$ and $\mathbf{M}$ in polar
form.  Recall that $Q_{11} = s\cos 2\vartheta$, $Q_{12} = s\sin
2\vartheta$ (cf. \Cref{sec:model}), and we can write $\mathbf{M} = r(\cos\phi,
\sin\phi)$ where $r = |\mathbf{M}|$ and $\phi$ is the angle between the $\Mvec$-vector and the $x$-axis. 
Using this parameterisation, the coupling energy density in \eqref{eq:energy} can be written as
\begin{equation}
  -c \Mvec^T\Qvec \Mvec = -cQ_{11}(M_1^2 - M_2^2) - 2cQ_{12}M_1M_2
  = -csr^2\cos 2(\vartheta - \phi).
  \label{eq:coupling_polar}
\end{equation}
The coupling energy density depends only on the \emph{relative angle}
$\vartheta - \phi$ between the nematic director and the magnetisation, and acts as a phase-locking potential between them.  
  
For $c = 0$, the Euler--Lagrange equations
for the angles $\vartheta$ and $\phi$ reduce to
\begin{equation}
  (l_1 s^2\,\vartheta')' = 0, \qquad (\xi l_2 r^2\,\phi')' = 0,
  \label{eq:angular_c0}
\end{equation}
expressing independently conserved angular fluxes. 
For $c \neq 0$, equations~\eqref{eq:angular_c0} are replaced by
\begin{equation}
  (4l_1 s^2\,\vartheta')' = 2csr^2\sin 2(\vartheta - \phi), \qquad
  (\xi l_2 r^2\,\phi')' = -2csr^2\sin 2(\vartheta - \phi),
  \label{eq:angular_c}
\end{equation}
where the right-hand sides are equal and opposite: the
two fields exchange angular momentum through the coupling torque $csr^2\sin
2(\vartheta - \phi)$.  This torque can support localised rotations, shifted
domain walls, and mixed-sign branches that cannot exist with $c=0$.
 
In summary, the separability of the $c = 0$ problem imposes two independent
conservation laws~\eqref{eq:angular_c0}, restricting the solution set to a
product structure.  Coupling breaks these conservation laws simultaneously,
replacing them with the single relation~\eqref{eq:angular_c} that connects both
fields.  These arguments, combined with orbit splitting and the loss of the product structure qualitatively explain the sharp increase in the number of critical points of \eqref{eq:energy} with $c\neq 0$.

\subsubsection{\texorpdfstring{$l=1$}{l=1} and \texorpdfstring{$c<1$}{c<1}} \label{sec:small-c}
For $c=0$, $l\geq 1.316$ is needed to stabilise the $OR$ solution. 
With $K=4\times 10^{-11}$ $N$ and $A=-0.172\times 10^6 $Nm$^{-2}$ \cite{Zumar2009}, this corresponds to a maximum channel width of $D = 1.33\times10^{-8}$m. OR solutions become unstable for larger channel widths.
Hence, we do not find a stable OR solution with $l=1$ and $c<1$. We find two stable full solutions and an unstable OR solution for $l=1$ and $c<1$.

\subsubsection{\texorpdfstring{$l=1$}{l=1} and \texorpdfstring{$c=3$}{c=3}}
The solution landscape of \eqref{eq:euler-lagrange} for $(l,c)=(1,3)$ is qualitatively the same as for $(l, c) = (0.2, 1)$. We find two full solution pairs: identified as $Stable\pm$, $Stable\mp$, $Stable+$ and $Stable-$, as for $l=0.2$ and $c=1$. These solutions are visually very similar to those in \Cref{fig:stable_l=0.2_c=1} (the only difference being $|\Qvec|:=\sqrt{Q_{11}^2+Q_{12}^2}$ and $|\Mvec|:=\sqrt{M_{1}^2+M_{2}^2}$ increase in the interior due to a larger value of $c$) and are hence, not plotted separately. There are nine unstable saddle points, of which there are three full saddle point solution pairs and three OR saddle points. All are visually similar to the saddle points plotted in \Cref{fig:full_saddle_points_l=0.2_c=1} and \Cref{fig:OR_saddle_points_l=0.2_c=1}, since they are continuations of them. 

We have the same number of stable, unstable, full and OR critical points with the same features, Morse index and connectivity for the $(l, c) = (0.2, 1)$ and $(l,c)=(1,3)$ cases respectively. By contrast, we only have a stable full solution pair and one unstable OR critical point for $l=1, c=1$. These observations support our hypothesis that increasing $c$ has the same impact as decreasing $l$: both increase the dominance of the nemato-magnetic coupling energy and this coupling proliferates the number of critical points of \eqref{eq:energy}.

\subsubsection{\texorpdfstring{$c=5$}{c=5}}

\begin{figure}[ht]
    \centering
    \begin{minipage}{0.38\textwidth}
        \centering
        \includegraphics[width=1.0\textwidth]{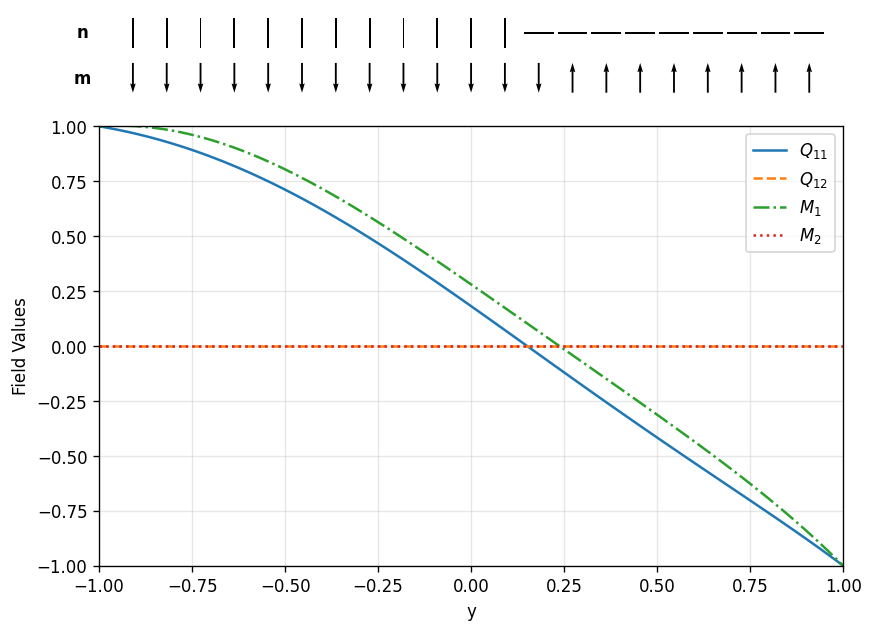}\\
        \ \textrm{$OR$ - stable}
    \end{minipage}
    \begin{minipage}{0.38\textwidth}
        \centering
        \includegraphics[width=1.0\textwidth]{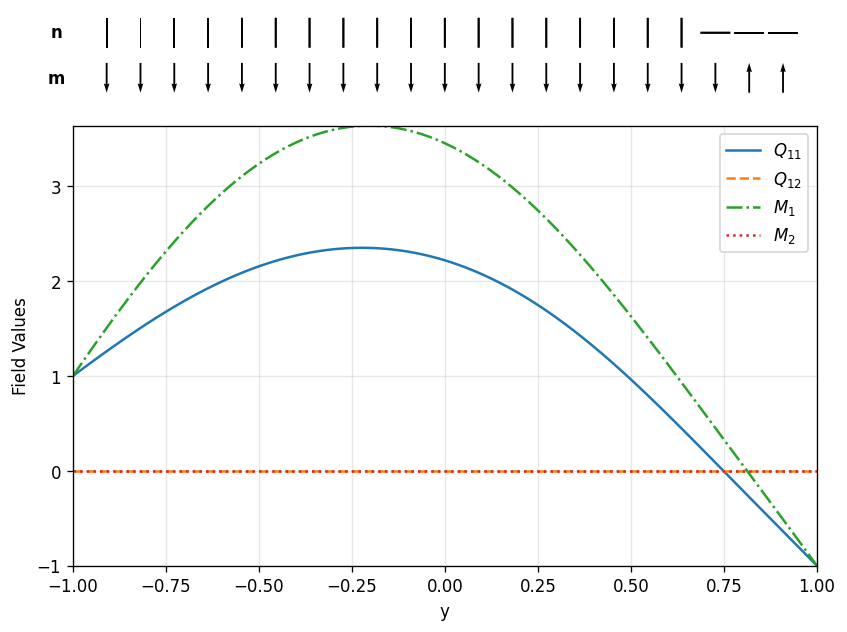}\\
        \ \textrm{$OR1$ - stable}
    \end{minipage}
    \begin{minipage}{0.38\textwidth}
        \centering
        \includegraphics[width=1.0\textwidth]{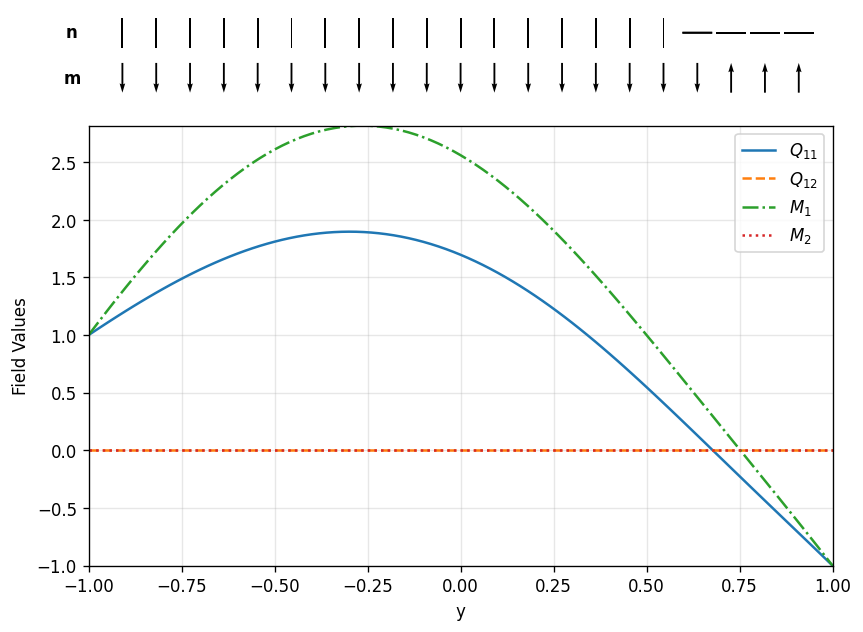}\\
        \ \textrm{$OR2$ - index-1 saddle}
    \end{minipage}
    \begin{minipage}{0.38\textwidth}
        \centering
        \includegraphics[width=1.0\textwidth]{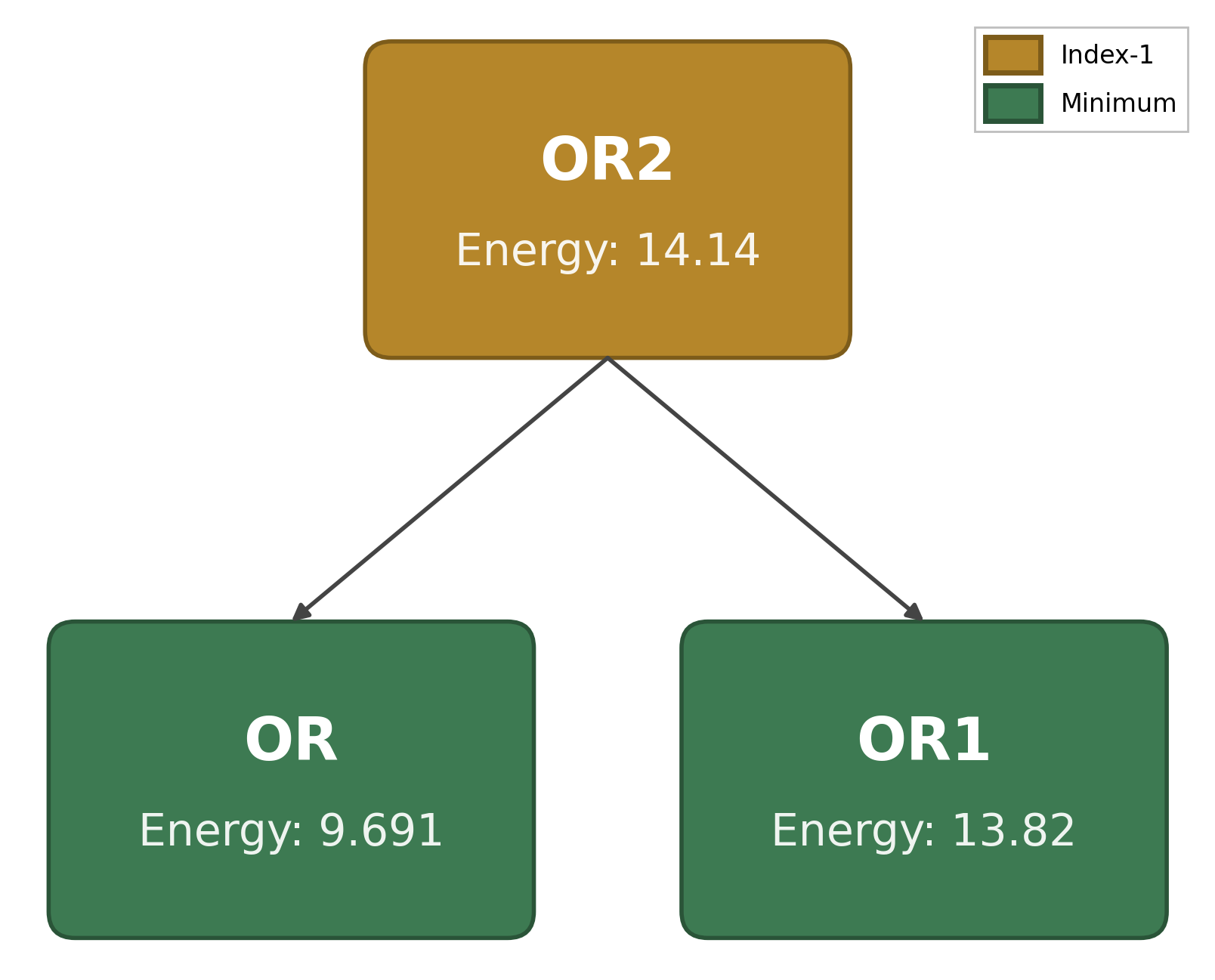}
    \end{minipage}
    \caption{The critical points of the energy \eqref{eq:energy} for $l=4.5$ and $c=5$ and the associated solution landscape.}
    \label{fig:saddle_points_l=4.5_c=5}
\end{figure}

\begin{figure} 
\centering
\includegraphics[width=0.75\textwidth]      {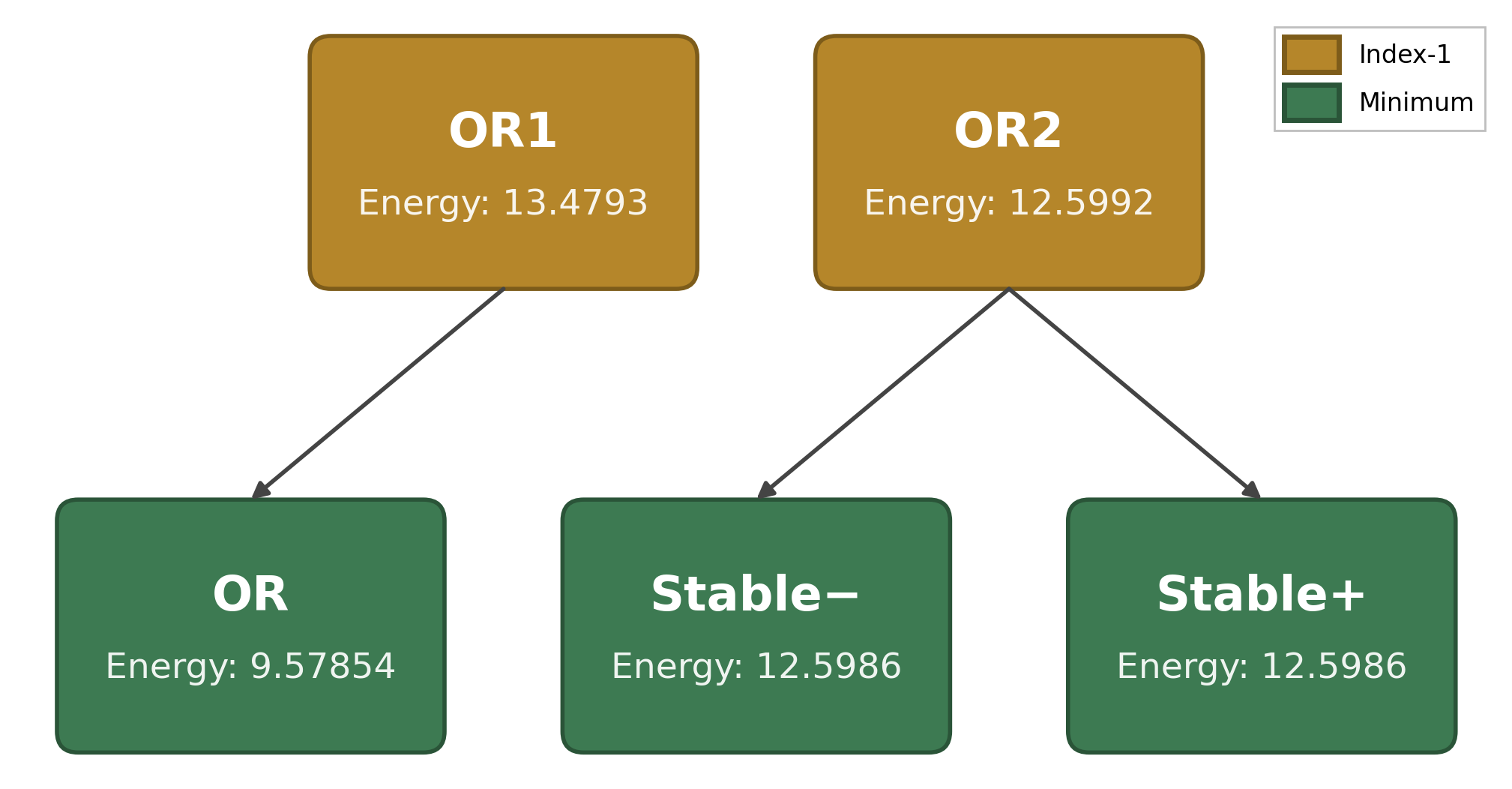}
    \caption{Solution landscapes for $l=4.45$ and $c=5$.}
    \label{fig:flow_diagrams_c=5}
\end{figure}

 To conclude this section, we now investigate the solution landscape for $c=5$ and different values of $l$. As mentioned in the Introduction, our parameter choices are academic in nature; they are not necessarily motivated by physical systems and in principle, $l$ and $c$ are tuneable material-dependent parameters.
For $l> 4.56$, there is a unique OR solution labelled as $OR$ and this solution persists for smaller values of $l$. We observe a new solution landscape for $l=4.5$, with two stable OR solutions and an index-$1$ $OR2$ saddle point (see \Cref{fig:saddle_points_l=4.5_c=5}). Namely, we can use the $OR2$ saddle point to connect the competing stable OR solutions (see \Cref{fig:flow_diagrams_c=5} row 1), which is an interesting observation. We do not find any full solution pairs in this parameter regime. All three OR solutions have one magnetic and one nematic domain wall in \Cref{fig:saddle_points_l=4.5_c=5}; the solution profiles and energies are similar for the stable $OR1$ and index-$1$ $OR2$ solution. In subsequent sections, we verify that this is not a numerical artefact but rather that one can generate families of OR solutions with minor differences in the locations of the domain walls. 

Decreasing $l$ further to $l=4.45$, we find a stable pair of full solutions, labelled as $Stable+$ and $Stable-$ respectively. 
The three OR solutions reported in \Cref{fig:saddle_points_l=4.5_c=5} exist for $l=4.45$. The associated solution landscape is plotted in \Cref{fig:flow_diagrams_c=5} row 2, for which $OR$ remains stable, $OR2$ remains index-$1$ and $OR1$ becomes an index-$1$ saddle point. We note that there is co-stability between $OR$ and the $Stable+$, $Stable-$ solutions, something not observed in our previous sections. However, there is no numerically computed pathway between the $OR$ and the stable full solutions, so that it is difficult to switch between the stable $OR$ and the stable full solutions. Moreover, the $OR$ is the global energy minimiser.

\begin{figure}
    \centering
    \includegraphics[width=0.7\textwidth]{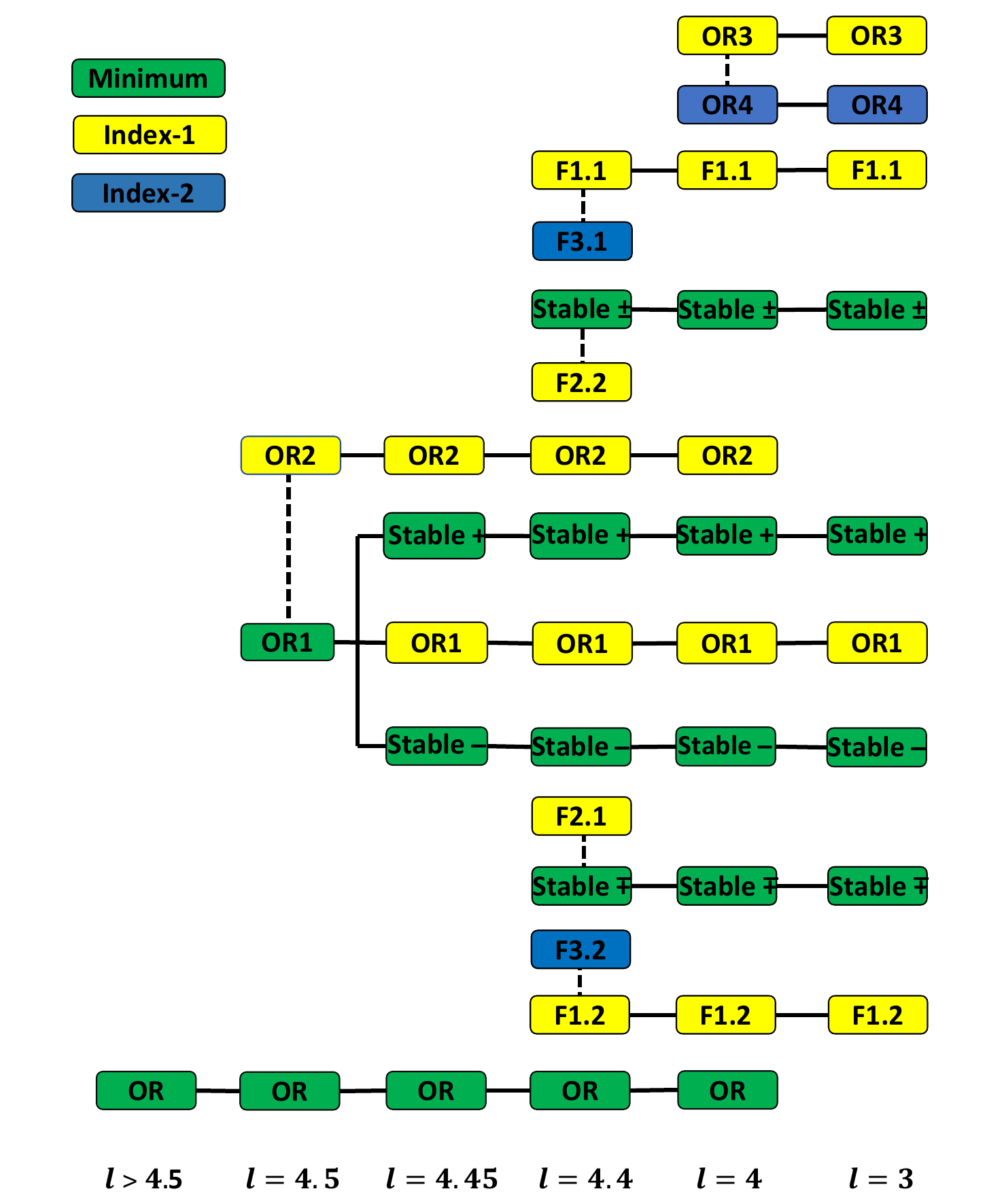}
    \caption{Bifurcation diagram for $c=5$. Solid lines indicate a continuous extension of a solution branch or emergence of a continuous pitchfork bifurcation, while dashed lines indicate a disconnected saddle node bifurcation.}
    \label{fig:bifurcation_c=5}
\end{figure}

In \Cref{fig:bifurcation_c=5}, we plot a bifurcation diagram to track how the solution landscape responds to decreasing $l$, for $c=5$.
For $l=4.4$ and $c=5$, we have five stable solutions: two full solution pairs as in \Cref{fig:stable_l=0.2_c=1}  
and the stable $OR$ solution. 
So we have co-stability between OR and full solutions, but no connection between them. We find eight unstable critical points: six full solutions and the index-$1$ saddle points, $OR1$ and $OR2$ respectively.  
For $l=4$ and $c=5$,  the unstable full solution pairs, $F2$ and $F3$, cease to exist and are replaced by two new OR solutions, $OR3$ (index-1) and $OR4$ (index-2 and the new parent state).
All other solutions persist in continuation from $l=4.4$ and their stability/Morse index is unchanged. We have five OR solutions for this parameter set, making OR solutions almost as prevalent as full solutions (of which there are six).  
Finally, for $l=3$ and $c=5$, the solution branches for $OR$ and $OR2$ cease to exist (they do not exist for $l<3.6$ approximately).  We no longer have co-stability between OR solutions and full solutions for $l \leq 3$ and $c=5$. 
All other solution branches persist and their Morse index is unchanged, so that $Stable\pm$, $Stable\mp$, $Stable+$ and $Stable-$ are the only numerically found stable solutions. In all cases, there may be higher-index solution branches or disconnected solution branches that are not captured by our methods.

The bifurcation diagram in Figure~\ref{fig:bifurcation_c=5} is an exploratory exercise with important lessons: (i) there exist parameter regimes characterised by solely OR solutions (with no full solutions); (ii) we can have co-stability between OR and full solutions; (iii) OR solutions, when unstable, usually connect stable solutions in a pair (so that $Q_{12}$ and $M_2$ can change sign across a nematic or a magnetic domain wall in an OR solution and this sign reversal dictates the pathway between the two stable full solutions in a given pair) but not stable solutions in different pairs; (iv) the unstable full solutions connect stable full solutions in different pairs; (v) the multiplicity of low-index OR solutions decreases as $c$ increases or as $l$ decreases, potentially due to the increasing energetic cost of domain walls with increasing $c$ or decreasing $l$; 
(vi) our numerical results suggest that there only exist two pairs of full stable solutions (with four degrees of freedom) for this model problem, for all $(l,c)$.

\subsection{Effects of negative \texorpdfstring{$c$}{c}}
In \cite{dalby-farrell-majumdar-xia}, and thus far in this paper, the authors only consider positive values of $c$. The results for positive $c$ can be translated directly to $c<0$ by means of the following proposition:

\begin{proposition}\label{lem:negative_c}
If $(\tilde{Q}_{11},\tilde{Q}_{12}, \tilde{M}_1,\tilde{M}_2)(y)$ is a solution of the Euler-Lagrange equations \eqref{eq:Q11_again}-\eqref{eq:M2_again} for $c>0$ and for given $l_1,l_2$ and $\xi$, subject to the boundary conditions in Table~\ref{table:BCs-ferronematics}, then $(-\tilde{Q}_{11},\tilde{Q}_{12}, -\tilde{M}_1,\tilde{M}_2)(-y)$  
is the corresponding solution to \eqref{eq:Q11_again}-\eqref{eq:M2_again} for $-c<0$, with the same set of $l_1,l_2$ and $\xi$, subject to the boundary conditions in Table~\ref{table:BCs-ferronematics}.
\end{proposition}

The proof follows by direct computation and is therefore omitted. We have computed the 1D ferronematic solution landscape for $(l,c) = (0.2, -1)$ and obtain the same number of stable solutions and saddle-point solutions of \eqref{eq:euler-lagrange} as in the case $(l,c) = (0.2,1)$. The difference is that 
$\nvec$ and $\mvec$ tend to be orthogonal to each other for $c<0$ whereas there is visible co-alignment between $\nvec$ and $\mvec$ for positive $c$, i.e., $(\nvec \cdot \mvec)^2 \approx 1$ almost everywhere in the channel interior, for stable full solution pairs. 

\subsection{Numerical Simulations with Realistic Parameter Values, Neumann Boundary Conditions and Two-dimensional Simulations}\label{sec:real}

In this section, we test the robustness of our 1D studies by working with more realistic parameter values, natural boundary conditions for $\Mvec$, and 2D analogues of some 1D solution landscapes. In all cases, we conclude that OR solutions are ubiquitous and their domain walls (that separate polydomains with constant $\nvec$ and $\mvec$) offer tremendous flexibility in connecting stable full solutions.

\begin{figure}[ht]
    \centering
    \begin{minipage}{0.38\textwidth}
        \centering
        \includegraphics[width=1.0\textwidth]{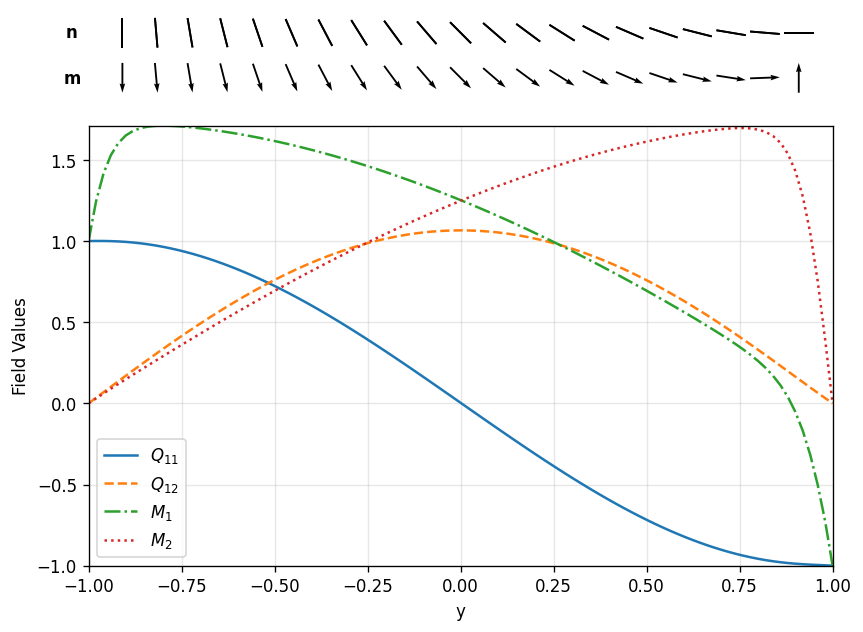}\\
        \ \textrm{$Stable+$}
    \end{minipage}
    \begin{minipage}{0.38\textwidth}
        \centering
        \includegraphics[width=1.0\textwidth]{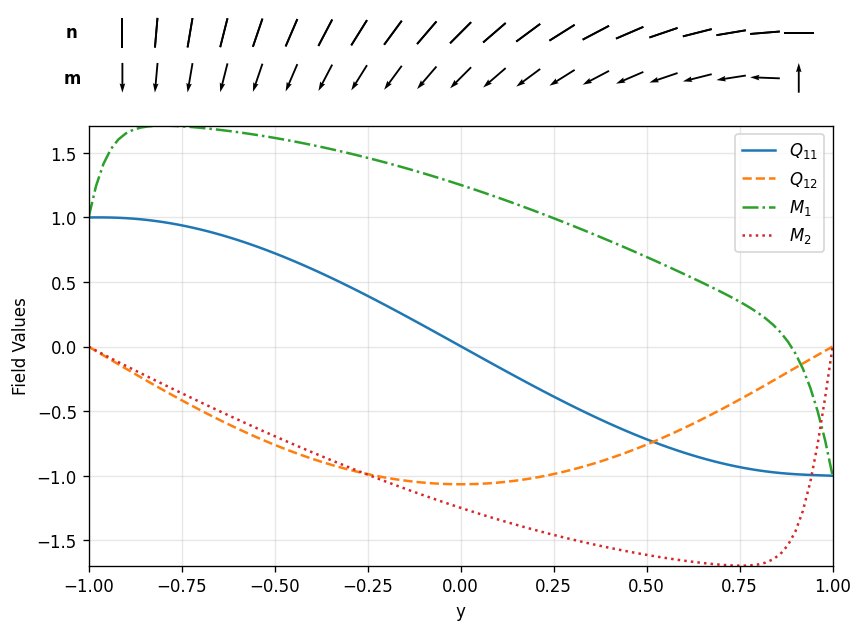}\\
        \ \textrm{$Stable-$}
    \end{minipage}
    \begin{minipage}{0.38\textwidth}
        \centering
        \includegraphics[width=1.0\textwidth]{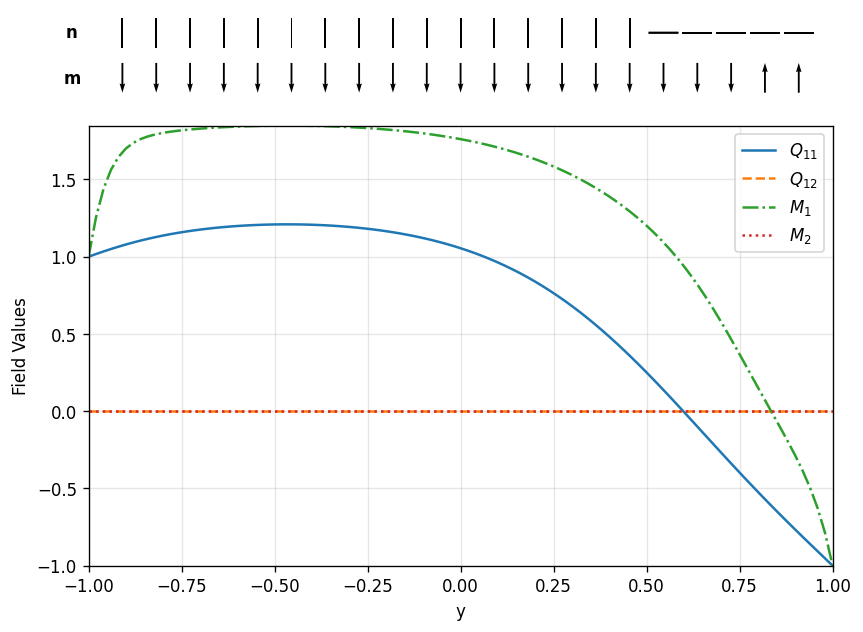}\\
        \ \textrm{$OR$ - index-2 saddle}
    \end{minipage}
    \begin{minipage}{0.38\textwidth}
        \centering
        \includegraphics[width=1.0\textwidth]{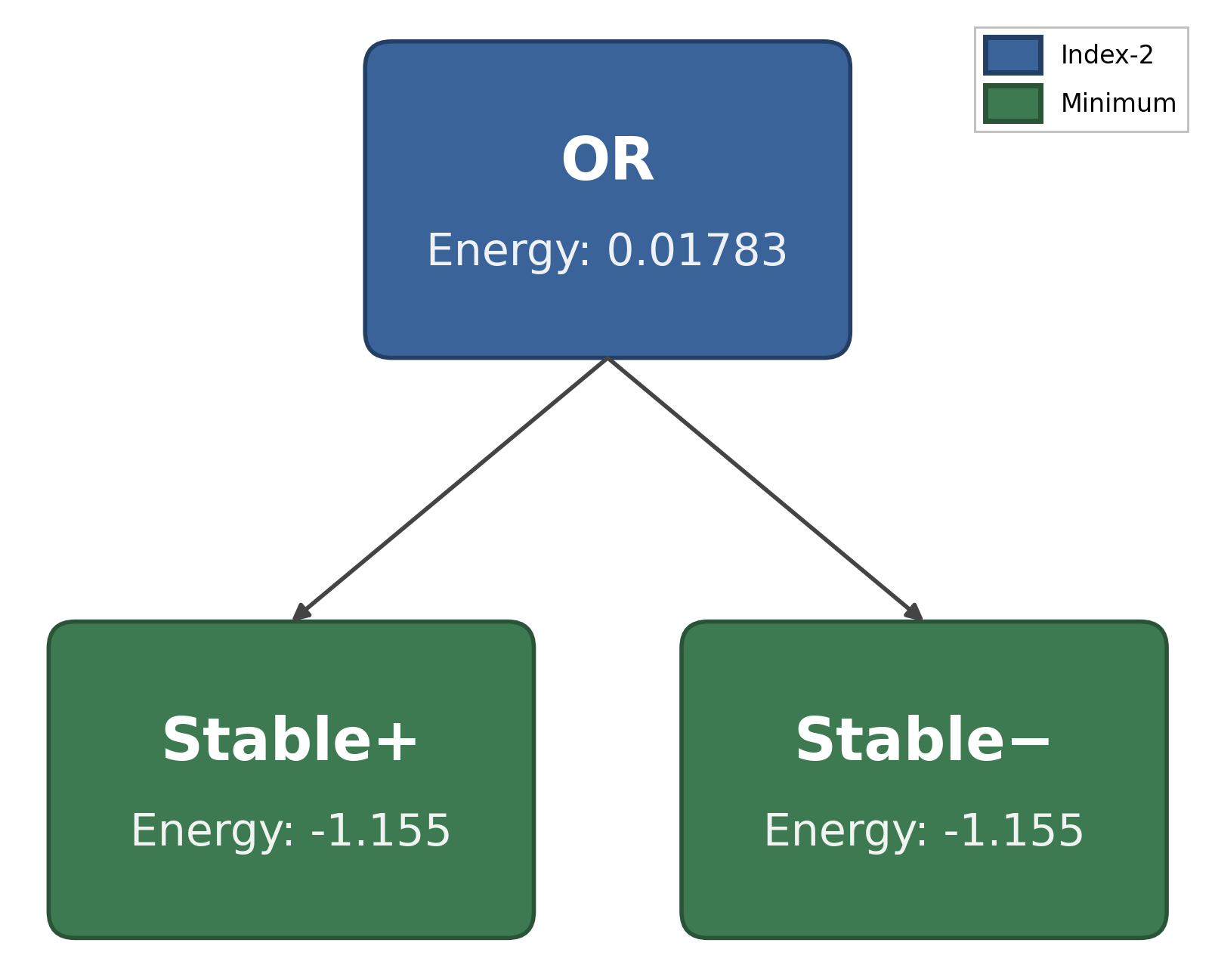}\\
        \ \textrm{$OR$ - index-2 saddle}
    \end{minipage}
    \caption{Critical points of \eqref{eq:energy} for $l_1=1, l_2=0.01$ and $c=\xi=1$ and associated solution landscape.}
    \label{fig:saddle_points_real_vals}
\end{figure}

We focus on the case $l_1=l_2:=l$ in this paper. The rescaled parameter $l_2$ is expected to be smaller than $l_1$ for dilute ferronematic systems. 
In Figure~\ref{fig:saddle_points_real_vals}, we plot two stable solutions and an unstable OR solution of \eqref{eq:euler-lagrange} with $l_1=1$, $l_2=0.01$, $c=\xi = 1$, subject to the Dirichlet boundary conditions for $\Qvec$ and $\Mvec$ as specified in Table~\ref{table:BCs-ferronematics}. Firstly, OR solutions still exist when $l_2$ is two orders of magnitude smaller than $l_1$ although the Morse index of the OR solution seems to be enhanced by a smaller value of $l_2$. This is simply because the energetic penalty of magnetic domain walls increases as $l_2$ decreases and this can have a further destabilising effect on the OR solutions. We plot two stable full solutions: the stable $Stable-$ solution and the $Stable +$ solution, connected by the index-$2$ $OR$ saddle point, exemplifying the importance of OR solutions even when $l_2 \ll l_1$.

\begin{figure}[ht]
    \centering
    \begin{minipage}{0.4\textwidth}
        \centering
        \includegraphics[width=1.0\textwidth]{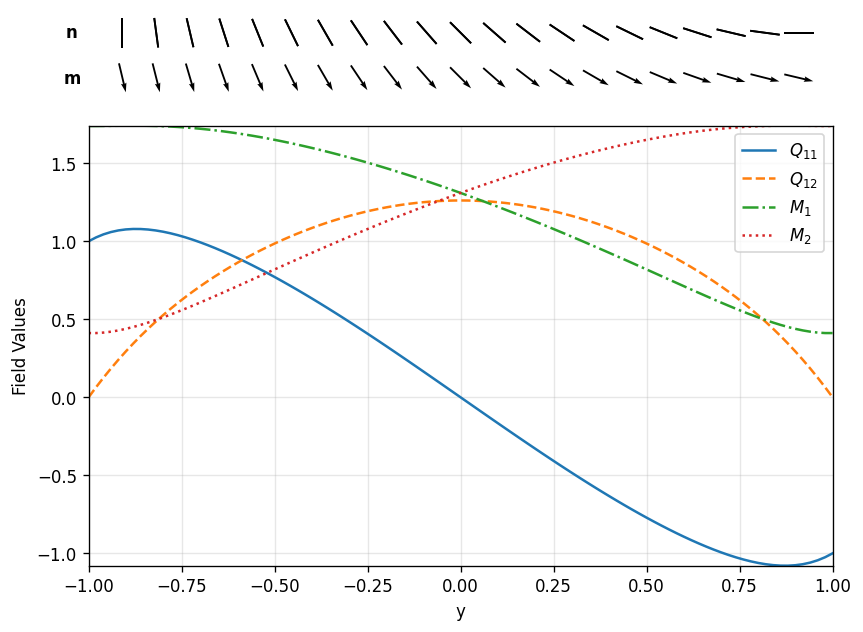}\\
        \ \textrm{$Stable+$}
    \end{minipage}
    \begin{minipage}{0.4\textwidth}
        \centering
        \includegraphics[width=1.0\textwidth]{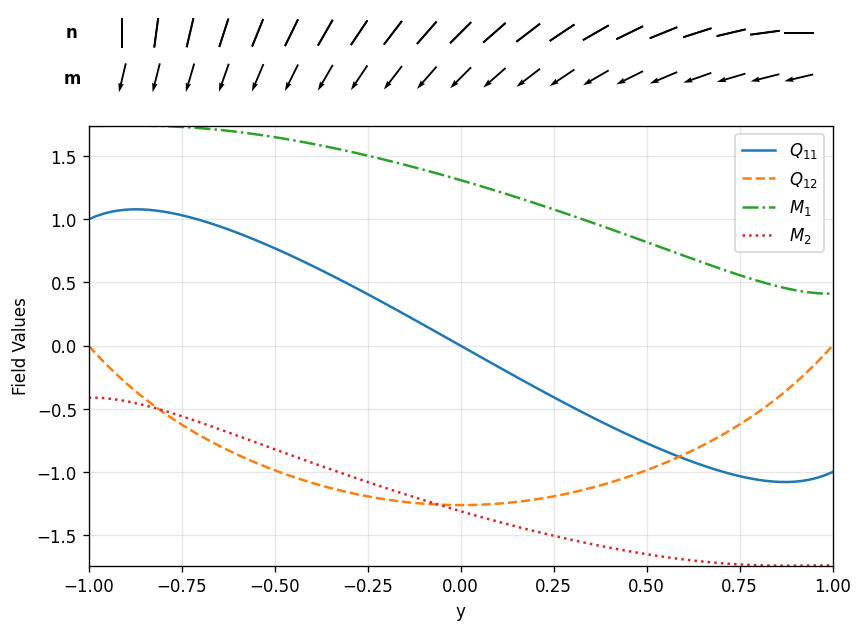}\\
        \ \textrm{$Stable-$}
    \end{minipage}
    \begin{minipage}{0.4\textwidth}
        \centering
        \includegraphics[width=1.0\textwidth]{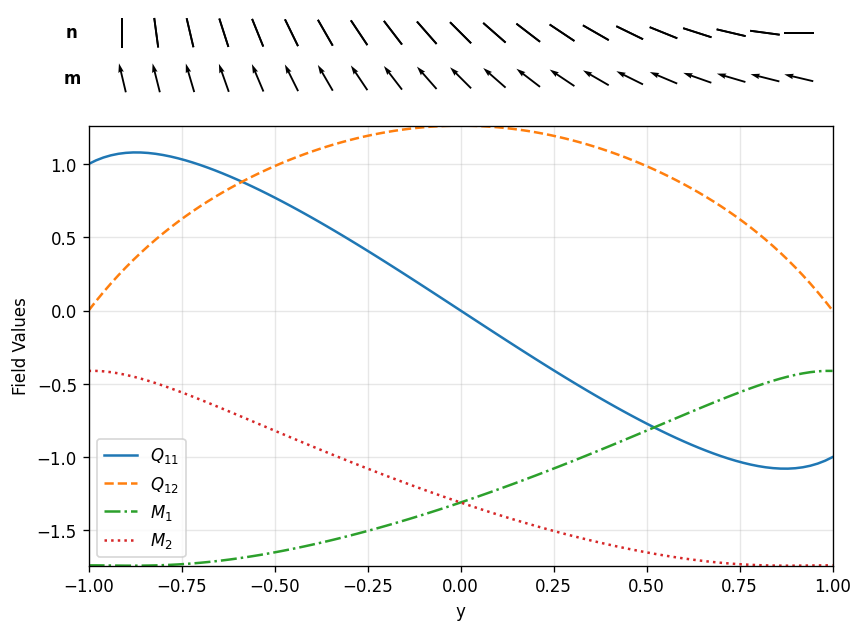}\\
        \ \textrm{$Stable\pm$}
    \end{minipage}
    \begin{minipage}{0.4\textwidth}
        \centering
        \includegraphics[width=1.0\textwidth]{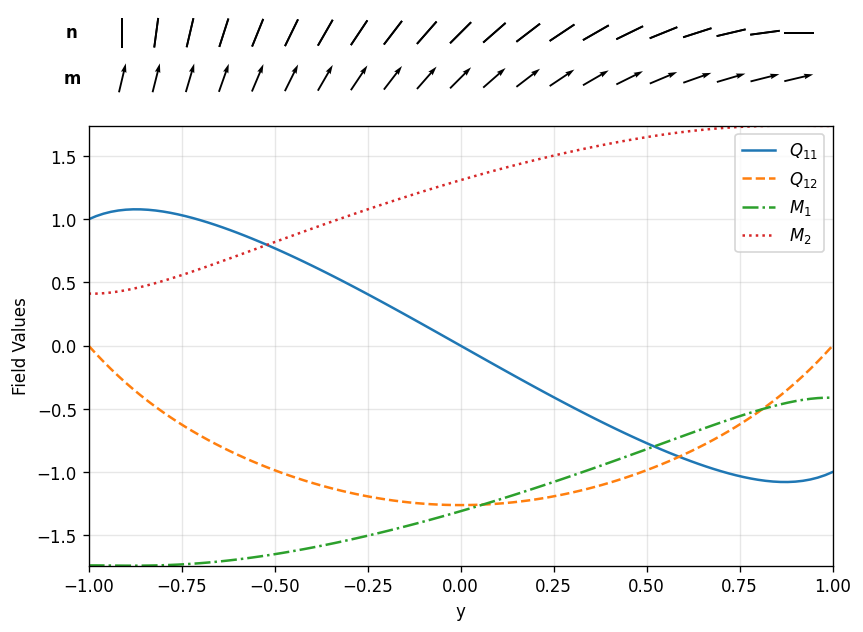}\\
        \ \textrm{$Stable\mp$}
    \end{minipage}
    \caption{Stable critical points of the energy \eqref{eq:energy} for $l_1=l_2=0.2$ and $c=\xi=1$, under zero-Neumann boundary conditions for $\Mvec$. }
    \label{fig:stable_points_neumann}
\end{figure}

\begin{figure}[ht]
    \centering
    \begin{minipage}{0.4\textwidth}
        \centering
        \includegraphics[width=1.0\textwidth]{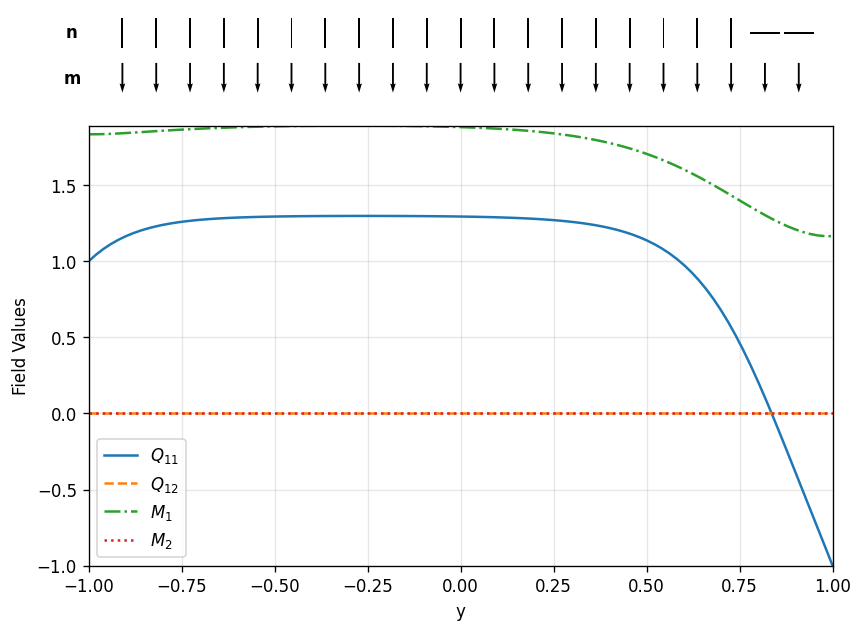}\\
        \ \textrm{$OR$ - index-1 saddle}
    \end{minipage}
    \begin{minipage}{0.4\textwidth}
        \centering
        \includegraphics[width=1.0\textwidth]{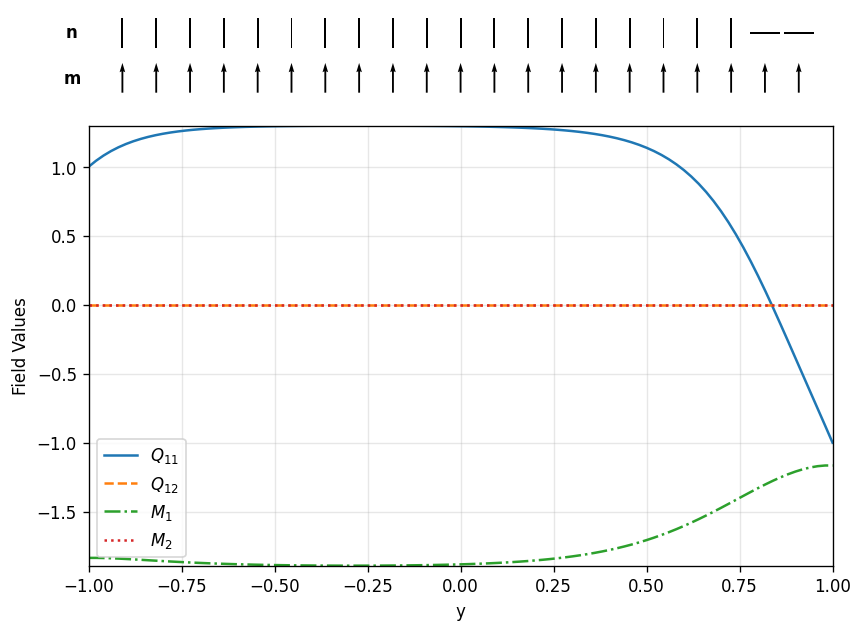}\\
        \ \textrm{$OR1$ - index-1 saddle}
    \end{minipage}
    \begin{minipage}{0.4\textwidth}
        \centering
        \includegraphics[width=1.0\textwidth]{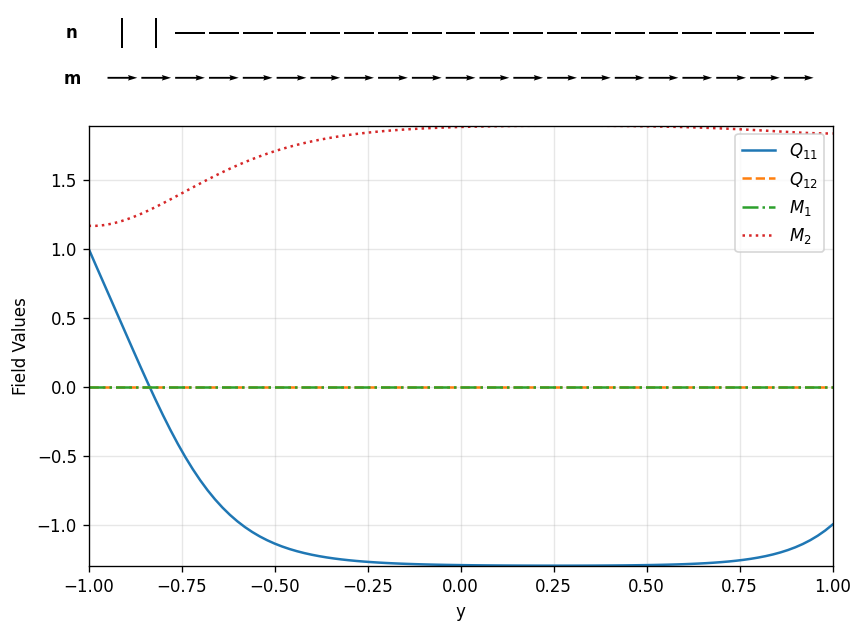}\\
        \ \textrm{$OR2$ - index-1 saddle}
    \end{minipage}
    \begin{minipage}{0.4\textwidth}
        \centering
        \includegraphics[width=1.0\textwidth]{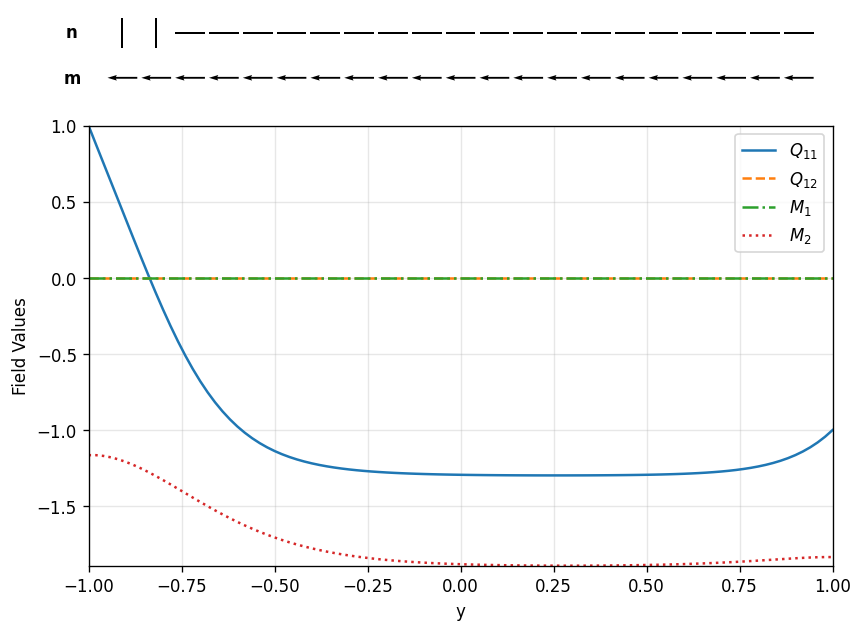}\\
        \ \textrm{$OR3$ - index-1 saddle}
    \end{minipage}
    \caption{Saddle points of the energy \eqref{eq:energy} for $l_1=l_2=0.2$ and $c=\xi=1$, under zero-Neumann boundary conditions for $\Mvec$. }
    \label{fig:saddle_points_neumann}
\end{figure}

As commented before, Dirichlet boundary conditions may not be physically realistic or realisable for $\Mvec$. Natural or Neumann boundary conditions are candidate boundary conditions for $\Mvec$ and hence, in \Cref{fig:stable_points_neumann} and in \Cref{fig:saddle_points_neumann}, we plot the stable critical points and some saddle points of \eqref{eq:energy} with $l_1 = l_2 = 0.2$, $c=\xi=1$, Dirichlet conditions for $\Qvec$ as in Table~\ref{table:BCs-ferronematics} and Neumann conditions for $\Mvec$; see below
\begin{equation}
\frac{\partial \Mvec}{\partial y} = 0 \quad \textrm{on $y=\pm 1$}.
\end{equation}
There are no boundary constraints on $\Mvec$ and any rotation in $\Mvec$ is tailored by the nemato-magnetic coupling and the rotation of $\nvec$.

\begin{figure} 
\centering
\includegraphics[width=0.75\textwidth]      {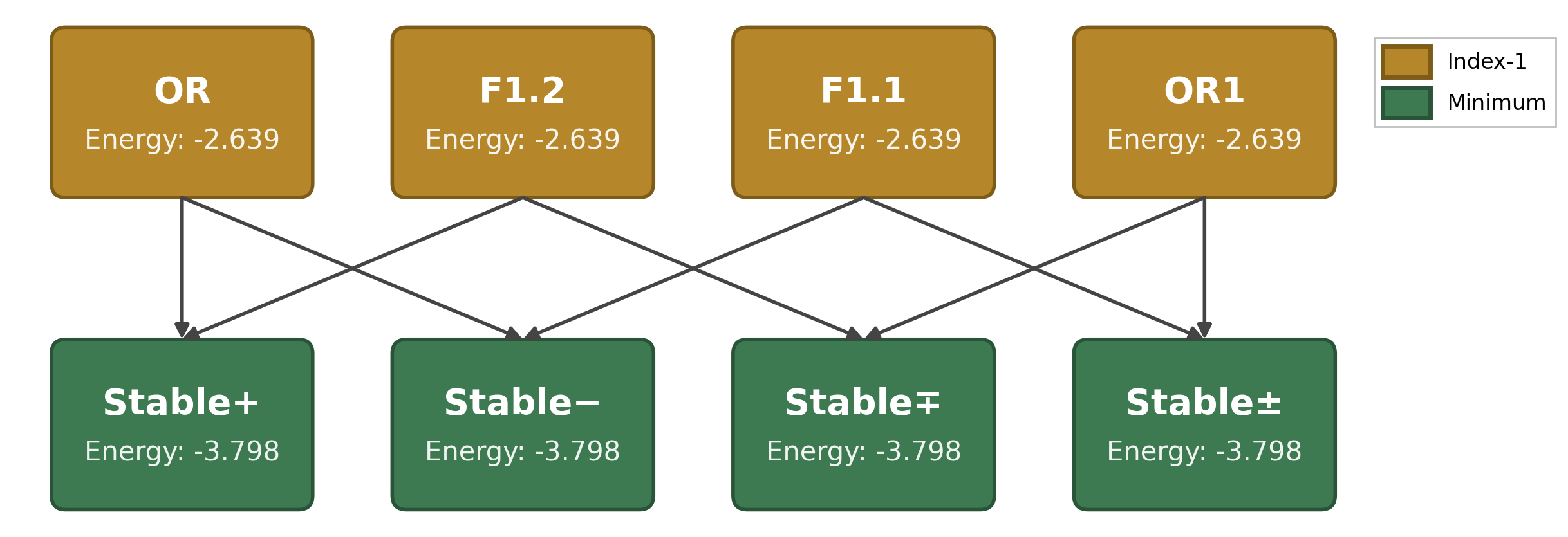}
    \caption{Solution landscapes for $l_1=l_2=0.2$ and $c=\xi=1$, under zero-Neumann boundary conditions for $\Mvec$. Note that all the stable critical points are energetically degenerate, as are the unstable critical points.} 
    \label{fig:flow_diagrams_neumann}
\end{figure}

We recover the four familiar full stable solutions, and four index-$1$ OR saddle points connecting these stable solutions (see \Cref{fig:flow_diagrams_neumann}). The OR solutions have a nematic domain wall with $\Qvec=0$, as necessitated by the Dirichlet conditions in Table~\ref{table:BCs-ferronematics}. The OR solutions have a constant $\mvec = \frac{\Mvec}{|\Mvec|}$ profile, i.e., $\mvec= (1,0)$ or $\mvec = (-1, 0)$, $\mvec = (0,1)$ or $\mvec = (0,-1)$ to ensure co-alignment with $\nvec$ in most of the channel interior. Of course, a constant $\mvec$ profile does not imply a constant $\Mvec$-profile, which is incompatible with the Euler-Lagrange equations in \eqref{eq:euler-lagrange}.
Thus, OR solutions still serve as transition states connecting competing stable solutions. In fact, searching for higher-index solutions, we find 4 index-2 OR solutions and index-$3$ unstable full solutions. Thus, Neumann boundary conditions for $\Mvec$ eliminate magnetic domain walls but increase the multiplicity of OR solutions, and their potential relevance in switching applications.

Next, we conduct a brief 2D study to test the robustness of our 1D study with respect to higher-dimensional perturbations. We consider the case $(l,c)= (4.5, 5)$ for which the 1D energy in \eqref{eq:energy} has two stable OR solutions, connected by an index-$1$ OR solution (recall the solution landscape in Figure~\ref{fig:saddle_points_l=4.5_c=5}). 

We take our non-dimensional 2D domain to be a rectangle: $[x,y]\in[-5,5]\times[-1,1]$, such that the length of the $x$-edge is $5$ times the width of the $y$-edge. This is consistent with our assumption that $D\ll L$. We impose periodic boundary conditions for $\Qvec$ and $\Mvec$ in the $x$-direction and work with the same Dirichlet conditions for $\Qvec$ and $\Mvec$ as stated in Table~\ref{table:BCs-ferronematics}. 

The 2D ferronematic energy differs from the 1D energy in \eqref{eq:energy} in the elastic energy terms, i.e. 
\begin{equation}
    \label{eq:2d-energy}
    \begin{aligned}
        F&(Q_{11}, Q_{12}, M_1, M_2)
    := \\
    &\int_{-5}^{5}\int^1_{-1} \Bigg\{\frac{l_1}{2}\left[\left( \frac{\partial Q_{11}}{\partial y}\right)^2 + \left(\frac{\partial Q_{11}}{\partial x}\right)^2+ \left(\frac{\partial Q_{12}}{\partial y}\right)^2 +  \left(\frac{\partial Q_{12}}{\partial x}\right)^2 \right] + \left( Q_{11}^2 + Q_{12}^2 - 1 \right)^2 \\
    & + \frac{\xi l_2}{2} \left[ \left(\frac{\partial M_1}{\partial y}\right)^2 + \left(\frac{\partial M_1}{\partial x}\right)^2+\left(\frac{\partial M_2}{\partial y}\right)^2 + \left(\frac{\partial M_2}{\partial x}\right)^2\right] + \frac{\xi}{4}\left(M_1^2 + M_2^2 - 1 \right)^2 \\
    & - cQ_{11}\left(M_1^2 - M_2^2 \right) - 2c Q_{12}M_1 M_2 \Bigg\} ~\mathrm{d}x\mathrm{d}y.
    \end{aligned}
\end{equation}
Remarkably, we have the $x$-invariant $OR$ and $OR1$ solutions as stable critical points of \eqref{eq:2d-energy}, connected by the $x$-invariant index-$1$ $OR2$ critical point of \eqref{eq:2d-energy}. In other words, the predictions of the simplified 1D model in \eqref{eq:energy} survive with higher-dimensional perturbations, at least for these parameter choices.
\begin{figure}[ht]
    \centering
    \begin{minipage}{0.5\textwidth}
        \centering
        \includegraphics[width=1.0\textwidth]{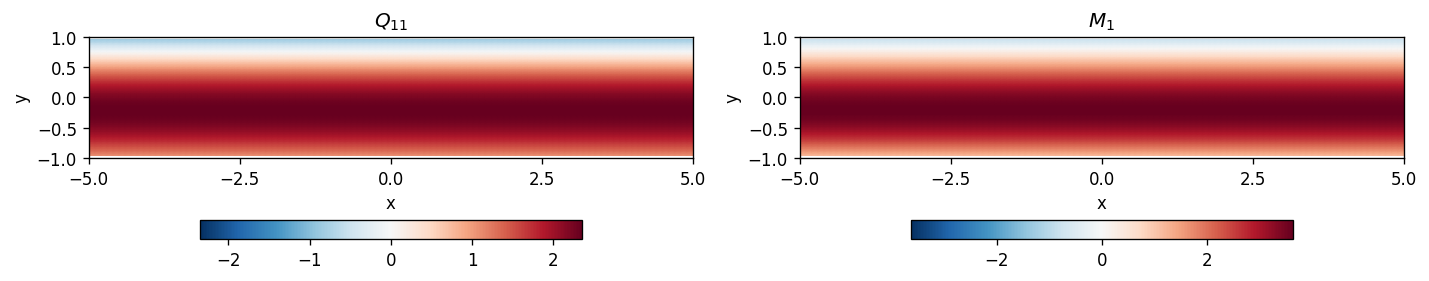}
    \end{minipage}
    \begin{minipage}{0.45\textwidth}
        \centering
        \includegraphics[width=1.0\textwidth]{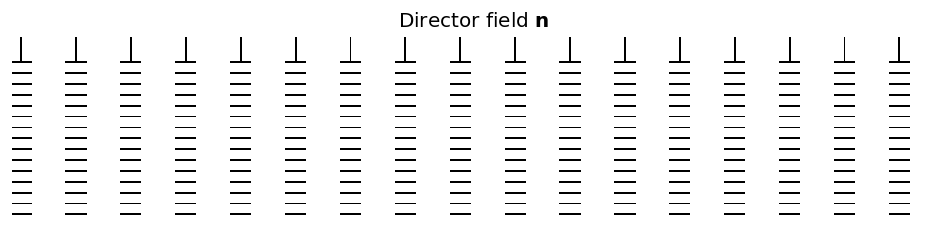}
    \end{minipage}
    \begin{minipage}{0.45\textwidth}
        \centering
        \includegraphics[width=1.0\textwidth]{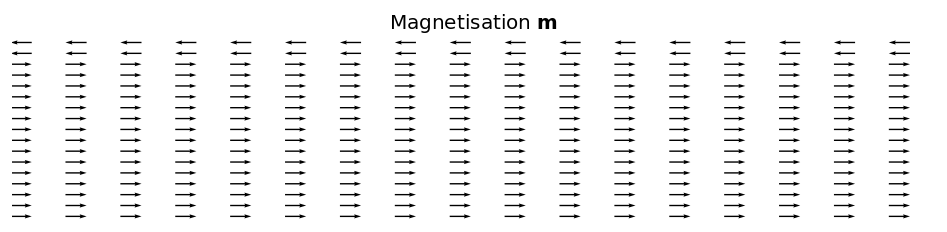}
    \end{minipage}
    \caption{2D stable $OR$ critical points of the energy \eqref{eq:2d-energy} for $l=4.5$ and $c=5$. The energy is $96.92$ and $OR$ is the energy minimiser.}
    \label{fig:OR_2d}
\end{figure}

\begin{figure}[ht]
    \centering
    \begin{minipage}{0.5\textwidth}
        \centering
        \includegraphics[width=1.0\textwidth]{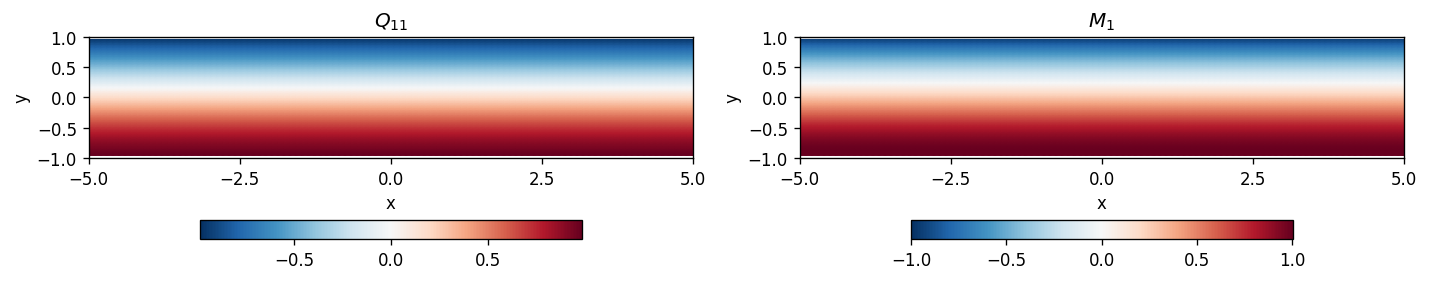}
    \end{minipage}
    \begin{minipage}{0.45\textwidth}
        \centering
        \includegraphics[width=1.0\textwidth]{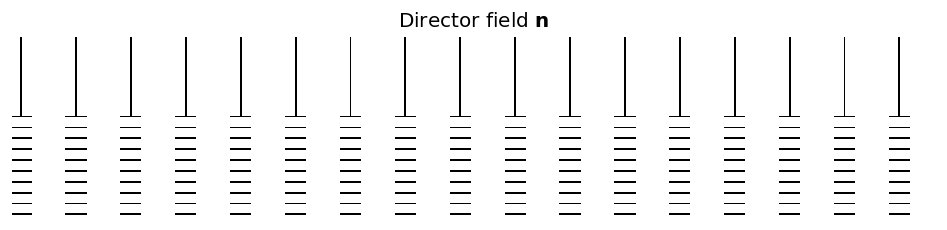}
    \end{minipage}
    \begin{minipage}{0.45\textwidth}
        \centering
        \includegraphics[width=1.0\textwidth]{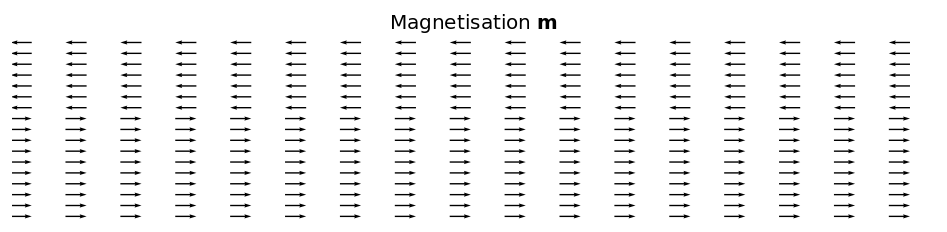}
    \end{minipage}
    \caption{2D stable $OR1$ critical points of the energy \eqref{eq:2d-energy} for $l=4.5$ and $c=5$. The energy is $138.1$.}
    \label{fig:OR1_2d_stable}
\end{figure}

\begin{figure}[ht]
    \centering
    \begin{minipage}{0.5\textwidth}
        \centering
        \includegraphics[width=1.0\textwidth]{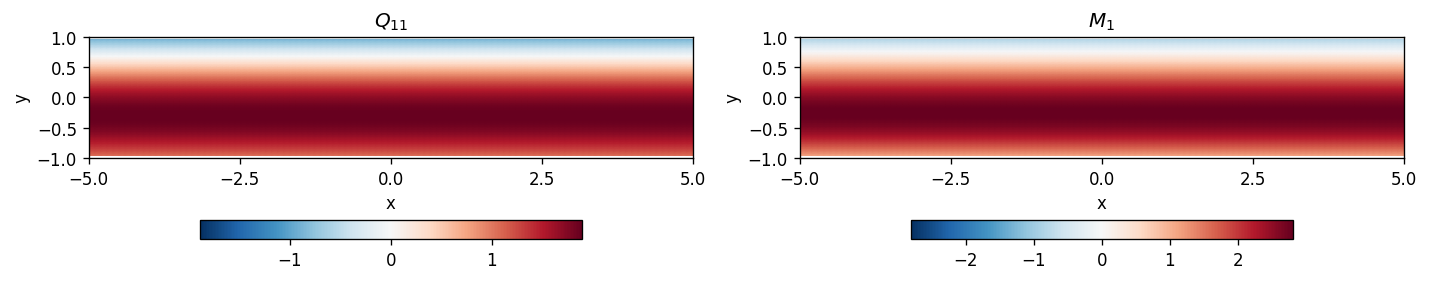}
    \end{minipage}
    \begin{minipage}{0.45\textwidth}
        \centering
        \includegraphics[width=1.0\textwidth]{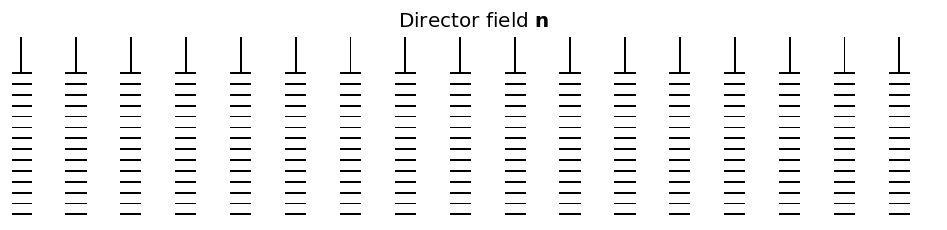}
    \end{minipage}
    \begin{minipage}{0.45\textwidth}
        \centering
        \includegraphics[width=1.0\textwidth]{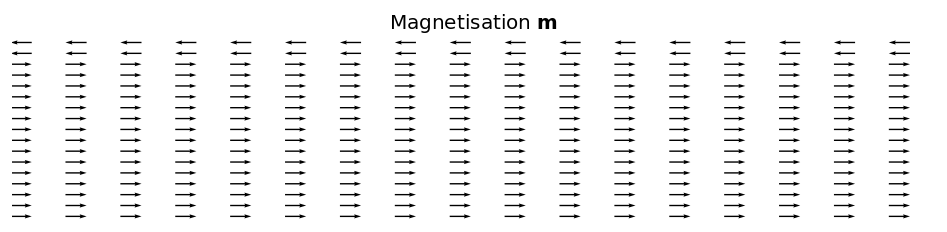}
    \end{minipage}
    \caption{2D index-$1$ $OR2$ critical points of the energy \eqref{eq:2d-energy} for $l=4.5$ and $c=5$. The energy is $141.4$.}
    \label{fig:OR_2d_saddle}
\end{figure}

 \section{Conclusions}\label{sec:conclusion}
In this paper, we perform an extensive exploration of solution landscapes of a 1D dilute ferronematic system, subject to Dirichlet boundary conditions for $\Qvec$ and $\Mvec$ in Table~\ref{table:BCs-ferronematics}, using the powerful HiOSD algorithm as described in \cite{yin2019high}. The ferronematic free energy is parameterised by four dimensionless parameters: $l_1, l_2, \xi$ and $c$ \cite{dalby-farrell-majumdar-xia}. For simplicity, we set $l_1 = l_2 := l$ and $\xi=1$, and study the impact of $l$ and $c$ on the solution landscapes, i.e., solutions of the Euler-Lagrange equations \eqref{eq:euler-lagrange}, their Morse indices and how the stable critical points are connected to one another by means of the unstable saddle points on the energy landscape. This can yield useful information about switchable multistable ferronematic devices.

Our numerical studies employ the reduced LdG framework for nematic phases, valid for certain 1D and 2D systems, and a 2D magnetisation vector. The rLdG $\Qvec$-tensor can be mapped to the full 3D $\Qvec_f$-tensor as in \eqref{eq:3dextrapolation}. We focus on two different types of solutions of \eqref{eq:euler-lagrange} in a simplified 1D setting: full solutions with four degrees of freedom, $(Q_{11}, Q_{12}, M_1, M_2)$; and OR solutions with two degrees of freedom such that $Q_{12}=M_2 \equiv 0$. We work with conflicting/frustrated Dirichlet conditions for $\Qvec$ and $\Mvec$ as outlined in Table~\ref{table:BCs-ferronematics}, which necessitate points with $\Qvec=0$ and $\Mvec=0$ in OR solutions. These points extend to 2D domain walls in the full 3D $\Qvec_f$ tensor, and are hence, referred to as nematic and magnetic domain walls. OR solutions are well-known as \emph{front solutions} since they separate polydomains with simple structures, i.e., nematic domain walls separate nematic polydomains such that $\nvec$ is constant (up to a sign) within each polydomain. Analogous remarks apply to magnetic polydomains. Whilst OR solutions arise in multiple contexts in materials science, we note that the nemato-magnetic coupling, embedded in the parameter $c$, proliferates the number of OR solutions and in some cases, gives rise to multiple stable OR solutions that differ in the location of nematic and magnetic domain walls.

In all cases, OR solutions play a pivotal role in the numerically computed solution landscapes. The OR solutions can be stable, act as transition states between stable solutions in a given pair or act as the parent state in the solution landscape. See for example Figure~\ref{fig:flow_diagrams_c=0}, for which the $OR$ solution is the index-$2$ parent state and can connect any pair of critical points (index-$1$ and index-$0$ critical points), and the energy barrier of a pathway mediated by the $OR$ solution is not substantially higher than those mediated by index-$1$ saddle points. In several examples, the OR solutions can act as transition states between stable full solutions in a given pair (see Figures \ref{fig:flow_diagram_l=0.2_c=1} and \ref{fig:flow_diagrams_c=5}). For $c=5$, there is co-stability between OR solutions and full solutions, so that OR solutions are both a selectable stable state and can also dictate selection dynamics, but there seems to be no pathway between a stable OR and a stable full solution. Perhaps our best discovery is for $c=5$ and $l=4.5$, where there are two stable OR solutions connected via an index-$1$ OR solution, and these are the only numerically computed critical points of the ferronematic free energy \eqref{eq:energy}. This is a useful and unexpected find as it illustrates that there are parameter regimes for which there are multistable ferronematic systems with multiple stable OR modes, i.e., we can shift the locations of domain walls and tailor them to targeted transport problems. Further work is recommended in this direction, i.e., can we truly design multistable ferronematic devices with experimentally observable polydomain structures?

The full solutions of \eqref{eq:euler-lagrange} have predictable structures - the stable full solutions usually describe smooth rotations of $\nvec$ and $\mvec$ to match the imposed boundary conditions. Any concentration of the rotation, either near the edges or near the centre, is associated with higher energies and a higher Morse index. Unstable full solutions can connect stable solutions in different families, e.g., $Stable +$ to $Stable \pm$, etc. Unstable full solutions need not have a fixed sign of $Q_{12}$ or $M_2$ for $-1 < y < 1$ and this sign-changing property dictates the pathway between stable full solutions in different families (which have a fixed sign of $Q_{12}$ and $M_2$ for $-1<y<1$).

Our numerical results show that \emph{fixing $c$ and decreasing $l$} has the same effect on the solution landscape as \emph{fixing $l$ and increasing $c$}, i.e., the number of stable and unstable critical points increases with decreasing $l$ and increasing $c$, as does the index of the parent state. 
For instance, the solution landscape in \Cref{fig:flow_diagram_l=0.2_c=1} is qualitatively repeated for three different choices of $(l,c)$: $(0.2,1)$, $(1,3)$ and $(0.2,-1)$. 
We speculate that the ratio $|\frac{l}{c} |$ 
 plays a pivotal role  and the repeating patterns are a consequence of the fact that the ratio $\frac{l}{c}$ is comparable for the parameter regimes, $l=0.2, c=1$ and $l=1, c=3$.

The number of critical points increases rapidly with decreasing $l$, e.g., there are at least $39$ solutions of \eqref{eq:euler-lagrange} when $(l,c)=(0.1, 1)$, and we expect approximately 98 solutions for $(l,c) = (0.01,1)$ (based on preliminary numerical tests). 
It remains an open question as to whether the large number of solutions (some of which appear visually similar, both for the full and OR solutions) is an artefact of the deterministic equations in \eqref{eq:euler-lagrange}. One could add noise terms to the right-hand side of \eqref{eq:euler-lagrange}; these noise terms could mimic modelling errors, material inhomogeneities or simply uncertainty in model parameters (see \cite{stochastic} for similar work in the rLdG framework for the planar bistable nematic device in \cite{tsakonas2007}) and then check how many solutions of \eqref{eq:euler-lagrange} survive with the addition of noise terms. For example, it is possible that there will be fewer stable and unstable OR solutions, or fewer unstable full solutions with the addition of noise terms in \eqref{eq:euler-lagrange}.

To conclude, we comment on the validity of the ferronematic free energy \eqref{eq:energy}. There are some open questions here: what are the admissible competing choices for the magnetic energy density; what are the different choices of the nemato-magnetic coupling energy and have we neglected physical effects stemming from the spontaneous magnetisation? We argue that this is a simplified model. 
The nemato-magnetic coupling energy in \eqref{eq:energy} is the simplest coupling energy that is consistent with frame indifference and material symmetry requirements. It can also be justified for dilute suspensions, using the rigorous homogenisation techniques adopted in \cite{canevarizarnescu}. However, in \cite{canevarizarnescu}, the authors average out the effects of the suspended nanoparticles in a NLC host, in terms of a constant field, $\mathbf{X}$, with corresponding homogenised energy, $\textrm{tr}\left(\Qvec \mathbf{X} \right)$.
In our case, $\mathbf{X}= \Mvec \Mvec^T$ but $\mathbf{X}$ is spatially varying in \eqref{eq:energy}. We do not have rigorous arguments to this effect but the suspended MNPs generate a spontaneous magnetisation and we speculate that the spontaneous magnetisation is a spatially inhomogeneous vector field. We also ignore any effects induced by MNP-MNP interactions and any potential out-of-plane magnetisation effects, although these effects are usually small for quasi 2D dilute ferronematic systems. 

We also briefly compare our ferronematic energy with the models employed in \cite{burylov}, \cite{smalyukh} and \cite{mertelj-2013-article}. These papers employ an Oseen-Frank approach to modelling NLCs, i.e., describe the nematic phase in terms of a unit-vector field, $\nvec$ instead of the rLdG $\Qvec$-tensor order parameter. An Oseen-Frank approach cannot capture OR solutions or domain walls, since $|\nvec|=1$ everywhere and nematic domain walls correspond to localised planar melting with $\Qvec=0$. In fact, an Oseen-Frank approach cannot capture all admissible unstable full solutions since there is no concept of a scalar order parameter in this framework. 
Further, these benchmark papers treat the spontaneous magnetisation as a spatially inhomogeneous field tailored by $\nvec$ but do not include any penalty terms involving $\nabla \Mvec$, i.e., $\Mvec$ is not treated as an independent order parameter in these papers. We, as in the papers \cite{burylov, mertelj-2013-article, smalyukh}, also treat $\Mvec$ to be a spatially inhomogeneous field and include all the terms in the magnetic energy from \cite{mertelj-2013-article} but also include an elastic energy density proportional to $|\nabla \Mvec|^2$, with stiffness constant $l_2$. Of course, $l_2$ should be small in practice, but we argue that an elastic penalty term is needed for spatially inhomogeneous $\Mvec$ fields and this should be carefully investigated in future work. In terms of agreement with experiments, polydomains and magnetic domain walls are reported in the pioneering ferronematic experiments in \cite{mertelj-2013-article} and these are precisely compatible with the OR solutions. Similar up-down domains have also been reported in the fascinating experiments described in \cite{smalyukh} and this further bolsters confidence in the importance of OR solutions for equilibrium and switching processes in novel ferronematic systems.

Our work is an informative example of a 1D soft matter system with two order parameters: a nematic tensor order parameter and a vector order parameter. The energy is quite generic and the concepts can be generalised to canonical systems with two coupled order parameters, and with varied boundary conditions in 2D and 3D. We expect the solution landscapes to be richer in 2D and 3D, i.e., more full solutions, more unstable OR solutions and greater choice in pathways (mediated by 2D and 3D saddle-points) between stable solutions with different energy barriers compared to the restricted 1D study in this paper. 
Hence, our work is certainly motivated by ferronematics and is applicable to certain types of ferronematic systems, but equally importantly, the simple free energy in \eqref{eq:energy} can help build overarching theoretical frameworks for archetypal soft matter systems.

\section*{Acknowledgements}
JD would like to thank Dr Yucen Han for helping him learn and understand the HiOSD algorithm, and for her valued feedback on the paper.
JD's PDRA position at the University of Strathclyde (during which time the work was completed) was funded by the EPSRC Additional Funding for Mathematical Sciences scheme. The authors gratefully acknowledge the anonymous referee's report from the paper's first submission, which pointed out the paper \cite{ZBD18b} and typical values for the parameters $c$ and $\xi$. A.M. gratefully acknowledges support from the Leverhulme Research Project Grant RPG-2021-401, a Royal Society of Edinburgh Network Grant and an Isaac Newton Institute Network Grant, EPSRC Grant EP/R014604/1.

\section*{Author contributions}
JD: Conceptualization, Formal analysis, Investigation, Software, Writing – original draft, Writing – review and editing. AM: Conceptualization, Analysis, Writing – review and editing.

\bibliographystyle{elsarticle-harv} 
\bibliography{main}

\end{document}